\documentclass[useAMS,usenatbib]{mn2e}
\usepackage{epsf}
\usepackage{amssymb}
\usepackage{graphicx}
\usepackage[usenames]{color}
\usepackage{fmtcount}

\newcommand {\bc}{\begin {center}}
\newcommand {\ec}{\end {center}}\newcommand {\be}{\begin {equation}}
\newcommand {\ee}{\end {equation}}
\newcommand {\disp}{\displaystyle}

\title[Gas Density Fluctuations in the Perseus Cluster]{Gas Density Fluctuations in the Perseus Cluster: Clumping Factor and Velocity Power Spectrum}
\author[Zhuravleva et al.]{I. Zhuravleva$^{1,2}$\thanks{zhur@stanford.edu}, E. Churazov$^{3,4}$, P. Ar{\'e}valo$^{5}$, A. A. Schekochihin$^{6,7}$, S. W. Allen$^{1,2,8}$, \newauthor A. C. Fabian$^{9}$, W. R. Forman$^{10}$, J. S. Sanders$^{11}$, A. Simionescu$^{12}$, R. Sunyaev$^{3,4}$, \newauthor A. Vikhlinin$^{10}$, N. Werner$^{1,2}$ \\
$^1$Kavli Institute for Particle Astrophysics and Cosmology, Stanford University, 452 Lomita Mall, Stanford, California 94305-4085, USA\\
$^2$Department of Physics, Stanford University, 382 Via Pueblo Mall,
Stanford, California 94305-4060, USA\\
$^3$Max Planck Institute for Astrophysics, Karl-Schwarzschild-Strasse 1, D-85741 Garching, Germany\\
$^4$Space Research Institute (IKI), Profsoyuznaya 84/32, Moscow
117997, Russia\\
$^5$Instituto de F{\'i}sica y Astronom{\'i}a, Facultad de Ciencias, Universidad de Valpara{\'i}so, Gran Bretana N 1111, Playa Ancha,
Valpara{\'i}so, Chile\\
$^6$Rudolf Peierls Centre for Theoretical Physics, University of Oxford, 1 Keble Rd, Oxford OX1 3NP, UK\\
$^7$Merton College, University of Oxford, Merton St, Oxford OX1 4JD, UK\\
$^8$SLAC National
Accelerator Laboratory, 2575 Sand Hill Road, Menlo Park, California 94025, USA\\
$^{9}$Institute of Astronomy, University of Cambridge, Madingley Road,
Cambridge CB3 0HA, UK\\
$^{10}$Harvard-Smithsonian Center for Astrophysics, 60 Garden Street, Cambridge, Massachusetts 02138, USA\\
$^{11}$Max-Planck-Institut f{\"u}r Extraterrestrische Physik,
Giessenbachstrasse 1, D-85748 Garching, Germany\\
$^{12}$Japan Aerospace Exploration Agency, 3-1-1 Yoshinodai, Sagamihara, Kanagawa 252-5210, Japan\\
}

\begin{document}

\date{Accepted .... Received ...}

\pagerange{\pageref{firstpage}--\pageref{lastpage}} \pubyear{2015}

\maketitle

\label{firstpage}

\begin{abstract} X-ray surface brightness fluctuations in the core of the Perseus Cluster are analyzed, using deep observations with the {\it Chandra} observatory. The amplitude of gas density fluctuations on different scales is measured in a set of radial annuli. It varies from $8$ to $12$ per cent on scales of $\sim 10-30$ kpc within radii of $ 30-160$ kpc from the cluster center and from $9$ to $7$ per cent on scales of $\sim20-30$ kpc in an outer, $160-220$ kpc annulus. Using a statistical linear relation between the observed amplitude of density fluctuations and predicted velocity, the characteristic velocity of gas motions on each scale is calculated. The typical amplitudes of the velocity outside the central $30$ kpc region are $90-140$ km/s on $\sim20-30$ kpc scales and $70-100$ km/s on smaller scales $\sim7-10$ kpc. The velocity power spectrum is consistent with cascade of turbulence and its slope is in a broad agreement with the slope for canonical Kolmogorov turbulence. The gas clumping factor estimated from the power spectrum of the density fluctuations is lower than $7-8$ per cent for radii $\sim 30-220$ kpc from the center, leading to a density bias of less than $3-4$ per cent in the cluster core. Uncertainties of the analysis are examined and discussed. Future measurements of the gas velocities with the {\it Astro-H}, {\it Athena} and {\it Smart-X} observatories will directly measure the gas density-velocity perturbation relation and further reduce systematic uncertainties in these quantities.
\end{abstract}
\begin{keywords}turbulence-methods: observational-methods: statistical-techniques: image processing-galaxies: clusters: intracluster medium-X-rays: galaxies: clusters
\end{keywords}

\section{Introduction}

The Perseus Cluster is the brightest galaxy cluster in the X-ray sky and has been attracting astronomers attention over the last 40 years. Being the X-ray brightest cluster, Perseus has led to many fundamental cluster discoveries, such as peaked cluster emission \citep{Fab74}, strong FeXXV and FeXXVI lines and thermal emission mechanism of the gas \citep{Mit76}, presence of bubbles of relativistic plasma \citep{Boe93}. The cluster was recently observed deeply with the {\it Chandra} \citep[see e.g.][and references therein]{Fab00,Fab11,Sch02,San07}, {\it XMM-Newton} \citep[see e.g.][and references therein]{Boe02,Chu03,Chu04,Mat13} and {\it Suzaku} \citep[see e.g.][and references therein]{Sim11,Ued13,Wer13,Tam14,Urb14} observatories, as well as by the previous-generation X-ray observatories \citep[see e.g.][]{For72,Mus81,Bra81,Arn94,Fab94,Fur01,Gas04}. These observations unveiled a variety of structures present in the hot gas of the cluster, reflecting richness and complexity of different physical processes occurring in the intracluster plasma. For example, an east-west asymmetry in the X-ray surface brightness (SB) \citep{Sch92}, aligned with a chain of bright galaxies, suggests an ongoing modest merger, which can induce sloshing of the gas and drive gas turbulence \citep{Chu03}, while the central few arcmin are dominated by AGN activity in a form of inflated bubbles of relativistic plasma \citep[see e.g.][]{Boe93,Chu00,Fab00} and weak shocks around them. Rising buoyantly and expanding, the bubbles uplift cooler X-ray gas from the core, producing filamentary structures seen in the optical, far-ultraviolet and soft X-ray, and excite internal waves, energy of which is transported to gas turbulence \citep{Chu01,Omm04,For07,Fab08,Can14,Hil14}.

The truly spectacular statistics accumulated by the {\it Chandra} observations of Perseus (over 85 million counts within 6 arcmin from the center, with 100s and 1000s counts per sq. arcsec) makes this data set an extremely powerful tool to probe  structures in the intracluster medium (ICM) on a range of spatial scales. Carefully modeling the Poisson noise, one can probe small-scale fluctuations, which we cannot see in the cluster images by visual inspection. Here, we present statistical analysis of these SB and density fluctuations, using the power spectrum (PS) statistics. We will examine the radial variations of the PS within the Perseus core ($220\times220$ kpc), while its energy dependence will be considered in our future works.

Statistical analysis of the X-ray SB and/or density, pressure fluctuations was done for two clusters so far: the Coma Cluster \citep{Sch04,Chu12} and AWM7 \citep{San12}. It was shown, for example, that the amplitude of density fluctuations in Coma ranges from $5$ to $10$ per cent on scales $30-500$ kpc, leading to nontrivial constraints on perturbations of the cluster gravitational potential, turbulence, entropy variations and metallicity. The fluctuations appear to be correlated, implying suppression of strong isotropic turbulence and conduction \citep{San13}.

The {\it Chandra} data on Perseus allow us to probe fluctuations on similar or even smaller scales, down to few kpc. The main questions we address here include: what is the amplitude of density fluctuations and how does it vary with the distance from the cluster center (Section \ref{sec:den_fluct}); what is the power spectrum of gas turbulent motions (Section \ref{sec:vel_ps}); how clumpy is the gas in the core (Section \ref{sec:clump}). To answer these questions, various instrumental effects (e.g. PSF variations, detector QE fluctuations), contribution of the unresolved point sources, subtraction of the Poisson noise as well as its fluctuations need to be treated carefully. We extensively investigate various systematic uncertainties of the analysis and discuss them in Section \ref{sec:uncert}. 

This paper is complementing our series of publications on SB and density fluctuations in the core of the Perseus Cluster. One application of the velocities of gas motions measured here is discussed in \citet{Zhu14b}. There we show that turbulent dissipation in the cluster core produces enough heat to balance locally radiative energy losses from the ICM at each radius. Therefore, turbulence may play an important role in AGN feebdack loop and be the key element in resolving the gas cooling flow problem \citep{Zhu14b}. In our next paper, we will discuss the nature of the fluctuations by measuring cross-spectra of fluctuations in different energy bands (Zhuravleva et al., 2015b, in prep.). Similar analysis for the Virgo and Coma clusters are in progress.

Throughout the paper we adopt $\Lambda$CDM cosmology with $\Omega_m=0.3$, $\Omega_{\Lambda}=0.7$ and $h=0.72$. At the redshift of Perseus, $z=0.01755$, the corresponding angular diameter distance is 71.5 Mpc and 1 arcmin corresponds to a physical scale 20.81 kpc. Position of the cluster center is taken from the NASA/IPAC Extragalactic Database\footnote{https://ned.ipac.caltech.edu/} and corresponds to the position of the central galaxy NGC1275. The total galactic HI column density\footnote{We use the mean value of the Galactic absorption within the field of the Perseus Cluster. Exclusion of soft photons ($0.5-1$ keV) from the SB analysis does not change the resulting amplitude of the SB and density fluctuations. Therefore, even if fluctuations of the Galactic absorption are present, their role in the cluster core is minor.} is 1.36$\cdot$10$^{21}$ cm$^{-2}$. Solar abundances of heavy elements are taken from \citet{And89}. We define the characteristic scale-dependent amplitude $A_{3D}(k)$ through the power spectrum $P_{3D}(k)$ as follows $A^2_{3D}(k)=\int 4\pi P_{3D}(k)k^2dk=4\pi P_{3D}(k) k^3$. A wavenumber $k$ is related to a spatial scale $l$ without a factor $2\pi$, i.e. $k=1/l$.

\section{Initial data processing}
\label{sec:init}
\begin{figure}
\begin{minipage}{0.49\textwidth}
\includegraphics[trim=0 150 0 100,width=1\textwidth]{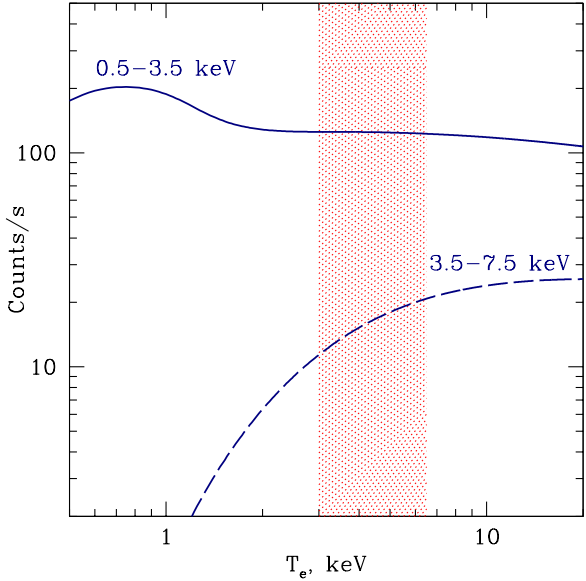}
\end{minipage}
\caption{Emissivity of an optically thin plasma as a function of temperature observed by the ACIS-S detector on {\it Chandra} in two energy bands: $0.5-3.5$ (solid) and $3.5-7.5$ (dashed) keV. Abundance of heavy elements is assumed 0.5 relative to Solar. Shaded region shows a range of gas temperatures in the core of the Perseus Cluster ($3-6.5$ keV). Notice, that for the $0.5-3.5$ keV band, the emissivity is almost independent of the temperature within this region. Therefore, the X-ray surface brightness in this shaded band $I_X$ is just a function of gas electron number density $n_e$, namely, $I_X\sim\int n_e^2 dl$, where $l$ is the length of the line of sight.   
\label{fig:fluxes}
}
\end{figure}

For our analysis we use public {\it Chandra} data with the
following ObsIDs: 3209, 4289, 4946, 4947, 4948, 4949, 4950, 4951,
4952, 4953, 6139, 6145, 6146, 11713, 11714, 11715, 11716, 12025,
12033, 12036, 12037. The initial data processing is done using
the latest calibration data and following the standard procedure
described in \citet{Vik05}. For the analysis, we choose $0.5-3.5$ keV energy band. In this band, the gas emissivity has a weak temperature dependence as shown in Fig. \ref{fig:fluxes}, and, therefore, the uncertainty associated with the conversion of PS of SB fluctuations to the PS of density fluctuations is reduced (see Section \ref{sec:den_fluct}).

\begin{figure*}
\includegraphics[trim=30 120 30 130,width=0.49\textwidth]{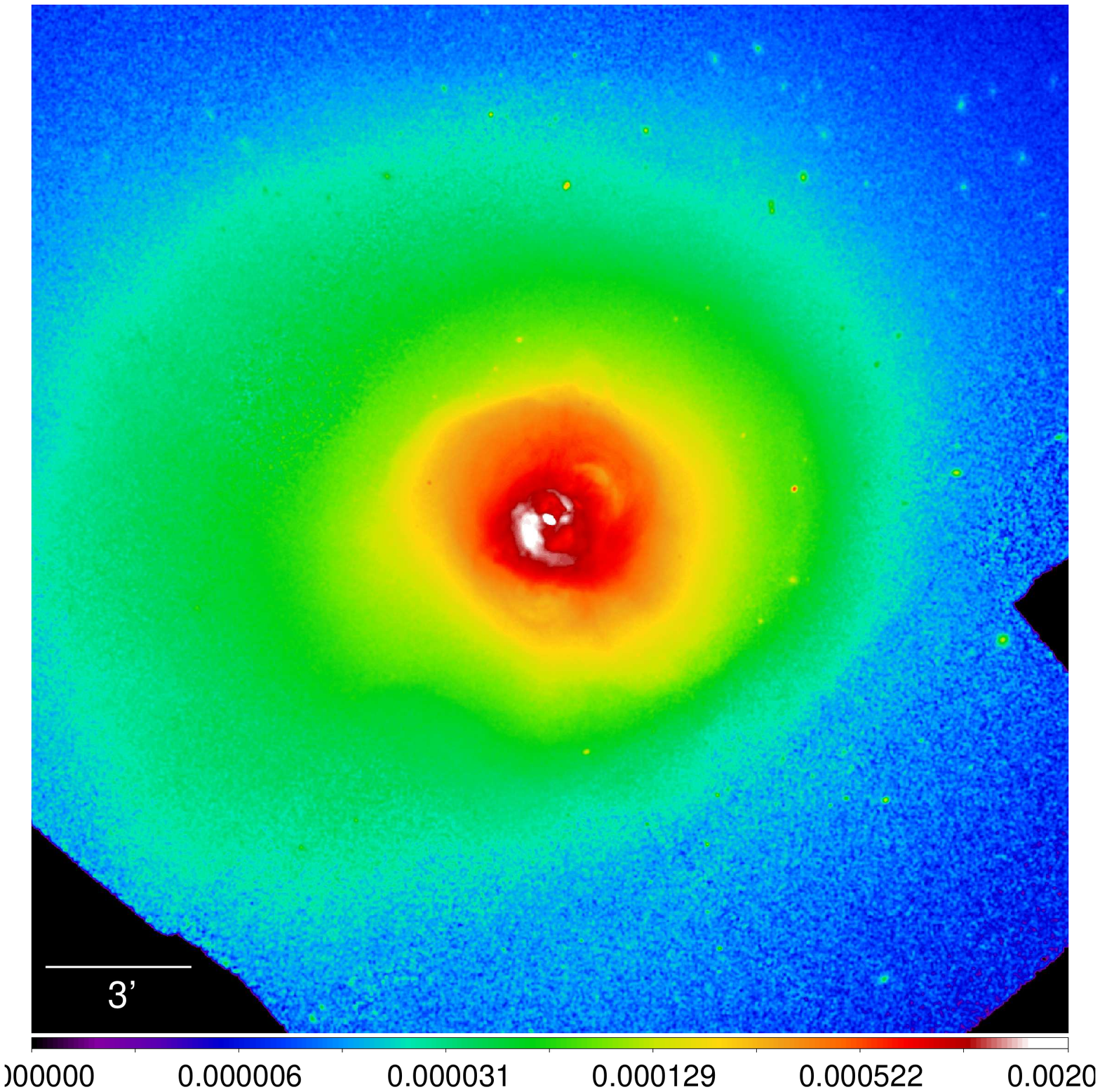}
\includegraphics[trim=30 120 30 130,width=0.49\textwidth]{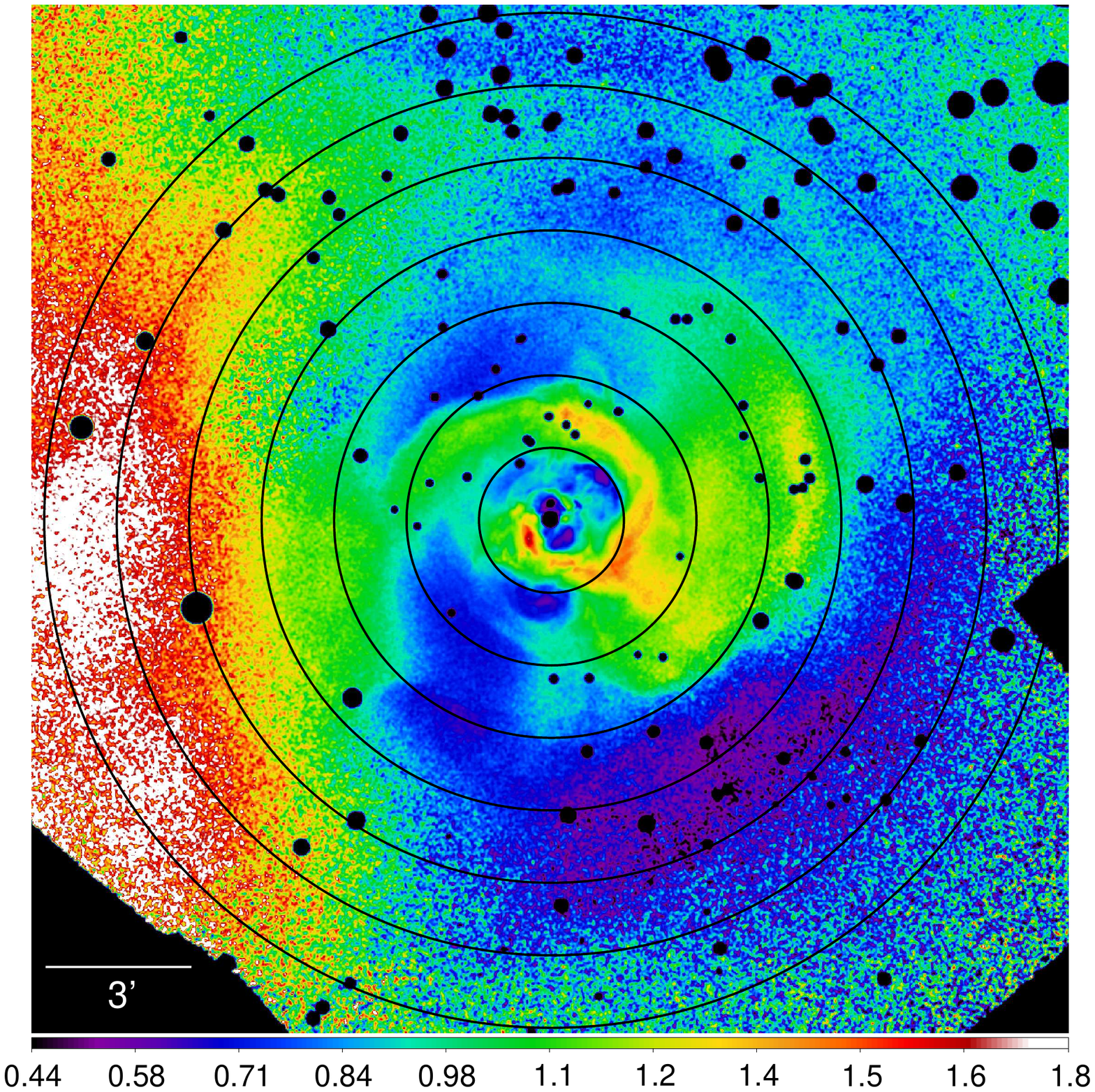}
\caption{{\bf Left:} {\it Chandra} mosaic image of the Perseus Cluster
  in the $0.5-3.5$ keV  band. The units are counts/s/pixel. The size of each pixel is 1 arcsec. {\bf
    Right:} residual image of the cluster (the initial image divided by
  the best-fitting spherically-symmetric $\beta-$model of the surface brightness), which emphasizes the surface
  brightness fluctuations present in the cluster. Point sources are
  excised from the image. Black circles show the set of annuli used in the analysis of the fluctuations. The width of each annulus is $1.5'$ ($\approx 30$ kpc). The
  outermost boundary is at a distance $10.5'$ ($\approx 220$ kpc) from the center. For display purposes, both images are lightly smoothed with a $3''$ Gaussian. 
\label{fig:im_pers}
}
\end{figure*}

\begin{figure*}
\includegraphics[trim=30 120 30 130,width=0.49\textwidth]{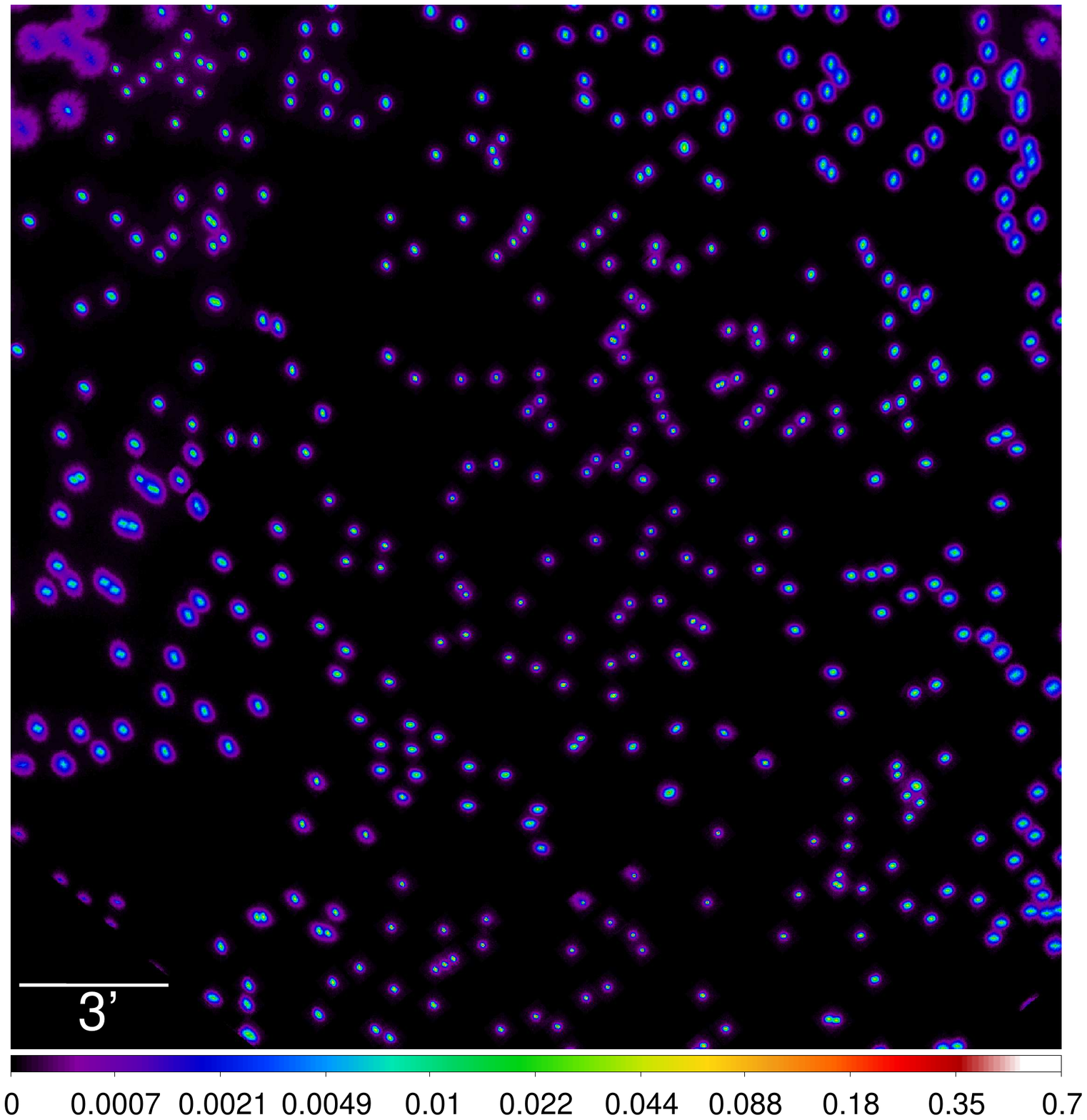}
\includegraphics[trim=30 120 30 130,width=0.49\textwidth]{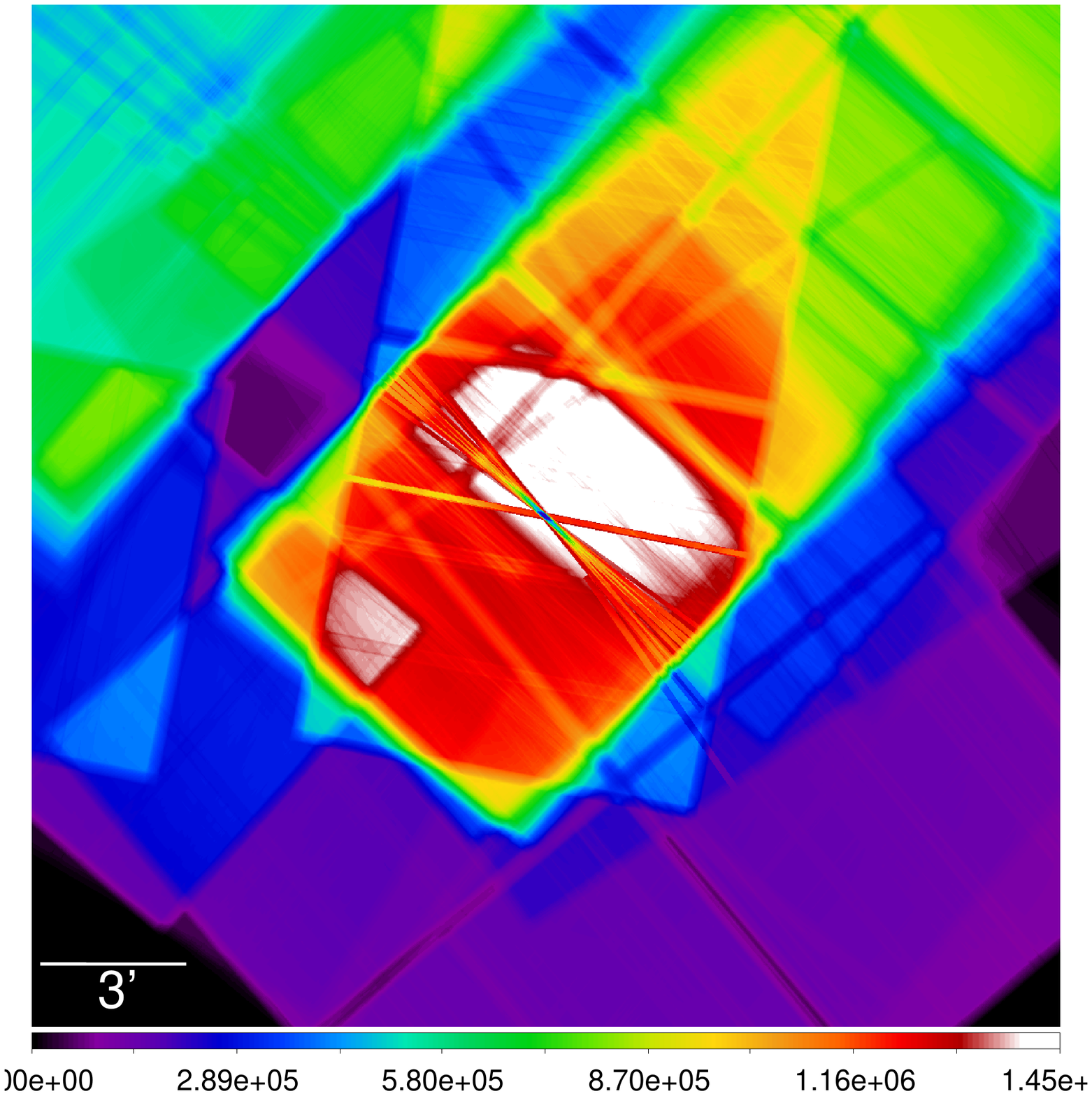}
\caption{{\bf Left:} simulated map of the combined {\it Chandra} PSF within the mosaic image of the Perseus Cluster. The random positions of individual
  PSFs are used (see Section \ref{sec:init} for details). {\bf Right:} the combined exposure map in seconds (lightly smoothed with a $3''$ Gaussian for display purposes).  
\label{fig:psf_exp}
}
\end{figure*}

Correcting for the exposure, vignetting effect and subtracting the background, a mosaic image of the Perseus Cluster is produced, which is shown in Fig. \ref{fig:im_pers}. The combined exposure map is shown in Fig. \ref{fig:psf_exp}. The total exposure of the cleaned data set is $\sim1.4$ Ms. Notice, that the
exposure coverage is inhomogeneous. It varies from few$\cdot
10^6$ s in the center down to few$\cdot 10^5$ s in some distal regions. The uncertainties associated with such inhomogeneous coverage are discussed in Section \ref{sec:uncert}.

The Perseus dataset includes observations with different offsets. For the analysis of SB fluctuations, it is essential to know the PSF at each point within the cluster image. We randomly generate positions, where the PSF is checked. The PSF at each position for each observation is determined from the {\it Chandra} PSF libraries \citep{Kar01}. By combining the PSF model images with weights proportional to exposure of each observation, the final PSF map is obtained. These maps are used to correct the PS of the SB fluctuations (Section \ref{sec:sb_fluct}). In order to illustrate the variations of the PSF shape within the cluster mosaic image, we show the PSF map divided by the exposure map in Fig. \ref{fig:psf_exp}. Notice a non-monotonic behavior of the PSF and its elongated (distorted) shape close to the edges of the image.

As the next step, we excise point sources from the image of the cluster. Using the wvdecomp tool, the point source candidates are found as peaks in the image with S/N $> 4$. In order to verify the significance of each source detection, we smooth heavily the image of the cluster with a $20''$ Gaussian, excluding the brightest point sources first. We then subtract the smoothed image from the initial image to remove the cluster contribution and calculate the point source fluxes in small regions around each point source candidate accounting for the combined PSF as $F_{\rm p.s.}=\sum I_{\rm p.s.}PSF/\sum PSF^2$, where $I_{\rm p.s.}$ refers to the residual image. The uncertainties on the fluxes are $\sigma_{F_{\rm p.s.}}=\sqrt{\sum C_iPSF^2}/\sum PSF^2$, where $C_i$ is the number of counts in the initial cluster image. Finally, sources with $F_{\rm p.s.}/\sigma_{F_{\rm p.s.}}>4$ are cut out from the images, using circles with the radius 1.5 times the 90 per cent radius of the combined PSF (see Fig. \ref{fig:im_pers}). Contribution of the unresolved point sources is estimated using sensitivity maps and Log N - Log S distribution of resolved sources. In case of the Perseus core, the contribution of unresolved point sources is negligible.

\section{Spherically-symmetric model}

\label{sec:model}
\begin{figure}
\begin{minipage}{0.49\textwidth}
\includegraphics[trim=20 155 30 90,width=1\textwidth]{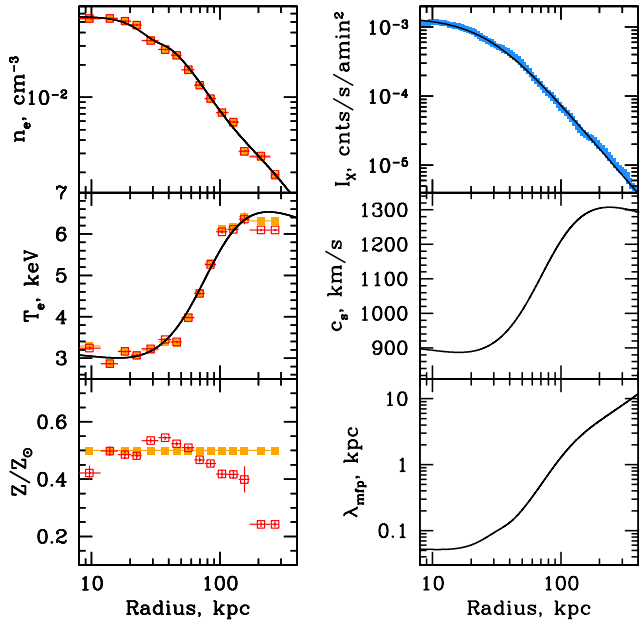}
\end{minipage}
\caption{Radial profiles of thermodynamic properties of the Perseus
  Cluster obtained from {\it Chandra} observations in the $0.6-9$ keV band. Orange (filled)
  and red (open) points with $1\sigma$ error bars on the left panels show the deprojected gas number electron
  density $n_e$, electron temperature $T_e$ and abundance of heavy
  elements relative to solar $Z/Z_{\odot}$, assuming the latter one
  is constant 0.5 relative to Solar or is a free
  parameter in the fitting spectral model of an optically thin plasma
  respectively. Analytic approximations of the observed profiles by smooth functions are shown with
  black solid curves. Blue points on the top right panel show the
  spherically-symmetric X-ray surface brightness profile obtained from the $0.5-3.5$ keV image, while the
  black curve shows its best-fitting $\beta-$model with the core radius
  $r_c=1.26$ arcmin and the slope $\beta=0.53$. The cluster sound speed
  $c_s$ and the mean free path $\lambda_{mfp}$ calculated from the
  approximations are shown in the middle and bottom panels on the right.
\label{fig:prof}
}
\end{figure}

The spherically-symmetric radial surface brightness profile of the Perseus is shown in Fig. \ref{fig:prof}. Fitting this profile with a $\beta-$model assuming photon counting noise only, gives the core radius $r_c=1.26$ arcmin and the slope of the model $\beta=0.53$ with negligible uncertainties. Allowing for 10 per cent systematic error on the SB in each annulus, the uncertainties become 0.1 and 0.01 for $r_c$ and $\beta$ respectively. Throughout the paper, this model is used as a default model of the unperturbed cluster. Sensitivity of our results to the choice of the model are discussed in Section \ref{sec:uncert}.  

The deprojected thermodynamic properties of the cluster, namely, the number electron density $n_e$ and the electron temperature $T_e$, are obtained from projected spectra using a procedure described in \cite{Chu03}. The deprojected spectra are fitted in the broad $0.6-9$ keV band using the XSPEC \citep{Smi01,Fos12} code and APEC plasma model based on ATOMDB version 2.0.1. The abundance of heavy elements is assumed to be either a) constant with radius 0.5 relative to Solar, or b) a free parameter in the plasma model. Fig. \ref{fig:prof} shows the deprojected data and their approximations with continuous, smooth functions. Using these approximations, the sound speed $c_s$ for an ideal monatomic gas is obtained as
\be
c_s=\disp\sqrt{\gamma\frac{k_BT_e}{\mu m_p}},
\ee
where $\gamma=5/3$ is the adiabatic index, $k_B$ is the Boltzmann constant, $\mu=0.61$ is the mean particle weight and $m_p$ is the proton mass. We also show the characteristic mean free path $\lambda_{\rm mfp}$ in the gas, which is numerically defined as \citep[e.g.][]{Sar88}
\be
\lambda_{\rm mfp}=23 {\rm \,kpc} \left(\frac{T_e}{10^8 {\rm \,K}}\right)^2\left(\frac{n_e}{10^{-3} {\rm \,cm^{-3}}}\right)^{-1}.
\ee  
Notice, that the mean free path is below kpc in central few tens of kpc and increases with distance, reaching $\sim 10$ kpc at distance 400 kpc from the center.

\section{Power spectrum of the surface brightness fluctuations}
\label{sec:sb_fluct}

X-ray SB of galaxy clusters is typically peaked towards centers and decreases rapidly with radius. This is especially true for relatively dynamically relaxed clusters like Perseus \citep{Man14}. Therefore, the PS of the initial SB will be dominated by this gradient, leading to a large power on small wavenumbers $k$ and contamination of the power on larger $k$. In order to avoid this, we first remove the global SB gradient and analyze fluctuating part of the SB only. In order to work with dimensionless units of the SB fluctuations, we divide the image of the Perseus Cluster by a simple spherically-symmetric $\beta-$model of the SB\footnote{One can also subtract the model, however in this case additional weights proportional to the global SB profile have to be introduced in order to treat the fluctuations at different radii equally. Clearly, the choice of this underlying model is arguable, but we will return to this problem in Section \ref{sec:uncert}.}.  Fig. \ref{fig:im_pers} shows the residual image of the SB fluctuations in Perseus. Notice a variety of structures of different sizes and morphologies present in the cluster, especially within the central $3$ arcmin. Let us quantify these structures statistically, by measuring their power as a function of length scale.

\begin{figure*}
\includegraphics[trim=10 180 30 70,width=0.49\textwidth]{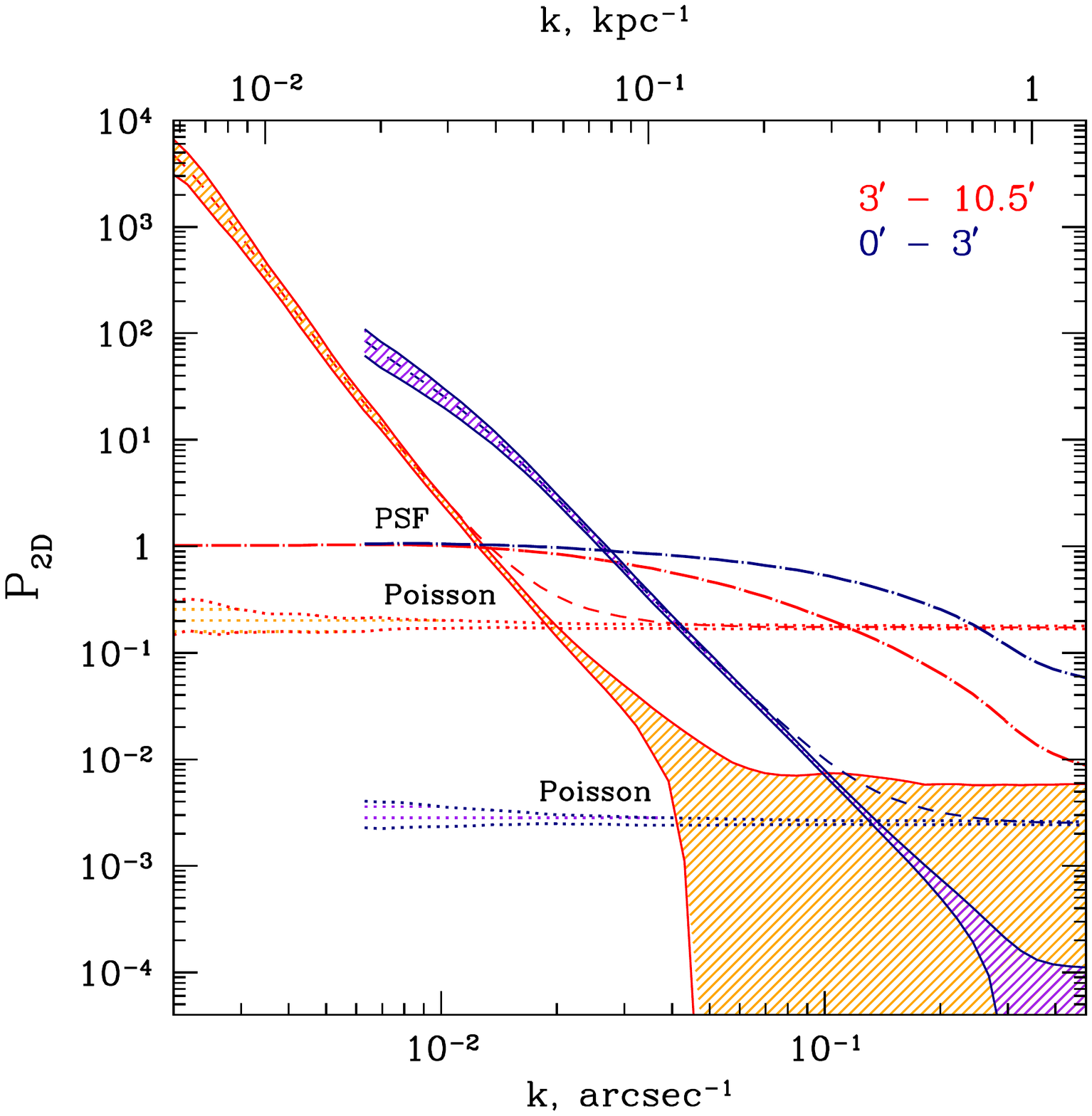}
\includegraphics[trim=10 180 30 70,width=0.49\textwidth]{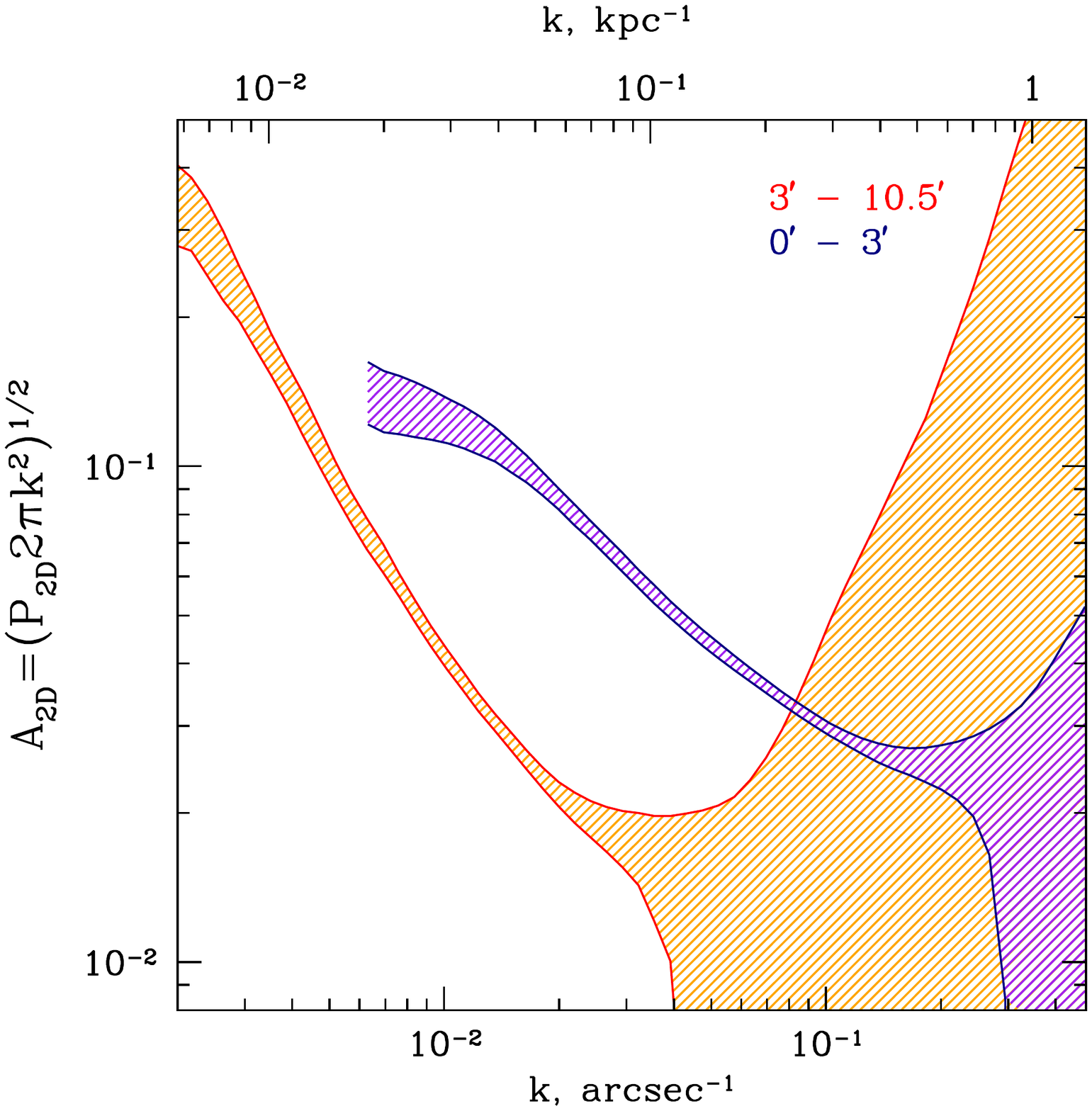}
\caption{{\bf Left:} power spectra (PS) of the X-ray surface
  brightness (SB) fluctuations in the Perseus Cluster in the central $3$ arcmin (purple) and
  in the annulus $3-10.5$ arcmin (orange). Dash curves: the initial
  PS. Dotted regions: PS of the Poisson noise. Dash-dotted curves: PS
  of the combined {\it Chandra} PSF in both analyzed regions. Hatched region with solid boundaries: PS of the SB fluctuations after subtraction of the Poisson noise. Both statistical ($1\sigma$) and stochastic uncertainties are taken into account. Notice, that very high statistics of counts accumulated by
  the 1.4 Ms observations allows us to probe fluctuations on a broad range
  of length scales, from $>$ 100 kpc down to few kpc (comparable with the mean free path). {\bf Right:} amplitude of the SB fluctuations in the same regions, obtained from the PS
  corrected for the PSF (see Section \ref{sec:sb_fluct}). Notice, that the amplitude of fluctuations varies with the radius. For these two regions a maximal difference by a factor of 4 is reached on $\sim 17$ kpc scale.
\label{fig:p2d_a2d}
}
\end{figure*}

We use a modified $\Delta-$variance method \citep{Are12} to calculate the PS of SB fluctuations. This method is tuned to cope with non-periodic data with gaps, the PS of which are smooth functions. Dashed curves in Fig. \ref{fig:p2d_a2d} show the PS of SB fluctuations in the central $3$ arcmin (central AGN is excluded) and in a broad annulus $3-10.5$ arcmin. The spectra decrease with wavenumber $k$ and reach constant value on small scales, where the signal is dominated by the Poisson noise.

Knowing the number of counts $n_{cnts}$ in each pixel in the cluster image in counts, the contribution of the Poisson noise is evaluated. A white noise has a flat spectrum; however, small deviations are still possible due to imperfections of the $\Delta-$variance method (e.g. due to filter behavior close to the edges of the image or near gaps, see \citet{Are12} for details). We made 100 realizations of the Poisson noise simply multiplying $\sqrt{n_{cnts}}$ by a random number from a normal distribution (with mean 0 and variance 1) in each pixel. For each realization, the PS of the Poisson image is calculated and, finally, the mean and the scatter of the resulting spectrum are obtained. Dotted horizontal regions in Fig. \ref{fig:p2d_a2d} show the level of the Poisson noise with the uncertainties in both considered regions. Subtracting the PS of the noise from the total (SB fluctuations+noise) PS, the PS of SB fluctuations is obtained (hatched region with solid boundaries in Fig. \ref{fig:p2d_a2d}).  The uncertainties reflect the photon counting noise at high frequencies ($1\sigma$ uncertainty is shown) and stochasticity of the signal at low frequencies. We do not separate these two types of uncertainties. Multiple realizations of the white noise PS $P_{\rm wn}$ estimate the scatter $\sigma_{p_{\rm wn}}$. These errors are then scaled to the measured PS of SB fluctuations $P$ by a simple multiplication 
\be
\sigma_p=\disp\sigma_{p_{\rm wn}}\frac{P}{P_{\rm wn}}.
\ee
Based on experiments with the whole process pipeline, we conservatively added 3 per cent systematic uncertainty to the scatter of the Poisson noise PS, which takes into account subtle numerical and instrumental effects. 

Finally, the PS of SB fluctuations have to be corrected for the {\it Chandra} PSF and for the contribution of the unresolved point sources on small scales. Performing multiple realizations of the PSF maps (see Section \ref{sec:init}), and each time calculating the PS of the map within the region of our interest, the mean spectrum of the PSF is obtained. Dash-dotted curves in Fig. \ref{fig:p2d_a2d} show the PSF PS in central $3$ arcmin and in the annulus $3 - 10.5$ arcmin. The deviations of the PSF shape from a $\delta-$function are reflected in a drop of power on the high-$k$ end of the spectrum. Dividing the PS of the SB fluctuations by the PSF power spectrum, the SB power suppressed by the PSF is recovered. 

The contribution of unresolved point sources is estimated by using the Perseus sensitivity map and by obtaining the shape and normalization of the Log N-Log S distribution of the resolved sources. In case of the Perseus Cluster, the contribution of the unresolved sources is negligible (within the cluster core).

After applying all corrections described above, the characteristic amplitude $A_{2D}(k)$ of the SB fluctuations is calculated, namely $A_{2D}(k)=\sqrt{P_{2D}(k)2\pi k^2}$, where $P_{2D}(k)$ is the measured PS. The amplitude is a more convenient characteristic since its units are the same of the variable in a real space. Fig. \ref{fig:p2d_a2d} shows the amplitude of SB fluctuations in the central $3$ arcmin region and in the outer annulus $3-10.5$ arcmin. High statistics of counts allows us to probe the SB fluctuations on a broad range of length scales - more than an order of magnitude, down to the scales comparable with the mean free path (see Fig. \ref{fig:prof}). The amplitude varies from $\sim 3-14$ per cent on scales $2-50$ kpc in the central $3$ arcmin to $\sim 2-33$ per cent on scales $\sim 14-170$ kpc in the $3-10.5$ arcmin region. The fact that the amplitudes in both regions differ at least by a factor of 3 over a broad range of scales shows that the amplitude of SB fluctuations in Perseus varies with radius. Therefore, instead of global characterization of the fluctuations in the cluster core, we consider below the fluctuations in a set of narrow radial annuli.

\section{Results}
\subsection{Power spectrum of density fluctuations}
\label{sec:den_fluct}
\begin{figure}
\begin{minipage}{0.49\textwidth}
\includegraphics[trim=0 150 0 75,width=1\textwidth]{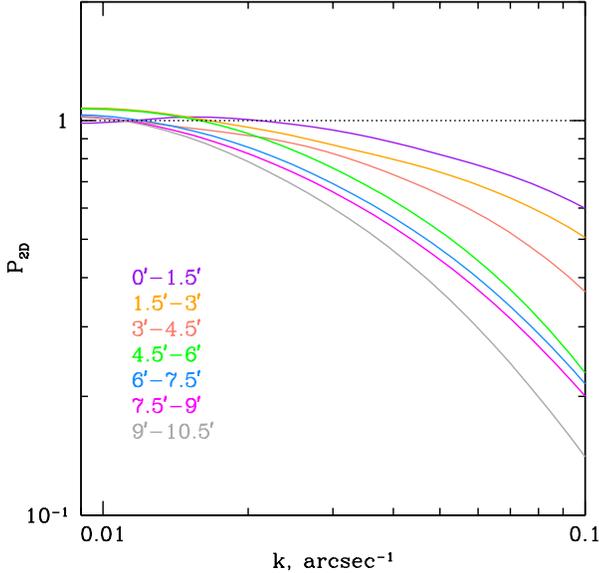}
\end{minipage}
\caption{Power spectrum (PS) of the combined {\it Chandra} PSF obtained in each annulus of the Perseus Cluster in our analysis (Fig. \ref{fig:im_pers}). The combined PSF accounts for the shape of the PSF at different offsets and exposure of each observation (see Section \ref{sec:init}). Each spectrum is averaged over 10 spectra of different realizations of the PSF map. Since the PSF is not a $\delta-$function, the PS deviate significantly from unity (dotted line), especially on scales smaller than 15 kpc ($k\approx0.02$ arcsec$^{-1}$).
\label{fig:psf}
}
\end{figure}

We divided the cluster into a set of radial annuli with a width $1.5'$ ($\approx 31$ kpc) (Fig. \ref{fig:im_pers}) and calculated the PS of SB fluctuations in each annulus, subtracting the Poisson noise and correcting for the unresolved point sources and the PSF. The latter one was done using PSF PS shown in Fig. \ref{fig:psf}. 

\begin{figure}
\begin{minipage}{0.49\textwidth}
\includegraphics[trim=0 250 0 200,width=1\textwidth]{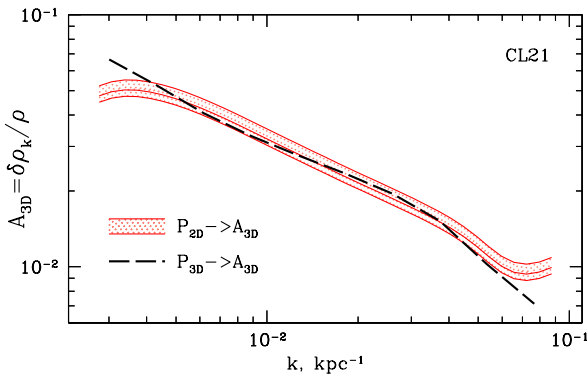}
\end{minipage}
\caption{Amplitude of density fluctuations in the inner 200 kpc (radius) region in the simulated relaxed
galaxy cluster (CL21 cluster with a total mass $M_{500c}=6.08 \cdot 10^{14} M_{\odot}$ at $r_{500c}=1215.2$ kpc, see \citet{Zhu14a} for details) obtained from the 3D density information (dash navyblue curve) and recovered
from the image of the X-ray surface brightness (hatched orange/red region).
The width of the hatched region reflects the uncertainty of
the recovered 3D amplitude of density fluctuations due to variations
of the converting geometrical factor with the projected radius. Notice
that the true (from the 3D data) amplitude of density fluctuations is within
the hatched region. This confirms that the observational procedure
of measuring the amplitude of density fluctuations from the X-ray
images recovers reasonably accurate the amplitude of the 3D density
fluctuations.
\label{fig:2dto3d_sim}
}
\end{figure}

The fact that the X-ray SB $I_X\sim \int n_e^2 dl$ for hot clusters with $T_e > 2$ keV (see Fig. \ref{fig:fluxes}) allows us to convert the 2D PS of SB fluctuations into the 3D PS of density fluctuations on scales smaller than the length of the line of sight. The procedure is described in \citet{Chu12}. Namely, knowing the global spherically-symmetric model of the gas emissivity, the conversion factor between the 3D and 2D power spectra at each line of sight $z$ is
\be
\frac{P_{2D}(k)}{P_{3D}(k)}\approx 4 \int |W(k_z)|^2dk_z,
\label{eq:p2dp3d}
\ee 
where $W(k_z)$ is the PS of the normalized emissivity distribution along the line of sight. 

We tested this procedure using high-resolution cosmological simulations of galaxy clusters \citep{Nag07,Nel14}. Using non-radiative simulations, we obtained the PS of density fluctuations directly from the simulated 3D data for a sample of 6 relatively relaxed clusters and recovered the PS from projected images of the X-ray SB (applying the same procedure we use for the observed X-ray images). The details of the sample and the analysis are described in \citet{Zhu14a}. Direct comparison of both PS (or amplitudes) confirms that the X-ray images can provide reasonably accurate measurements of the PS of density fluctuations. Fig. \ref{fig:2dto3d_sim} shows an example of one of the clusters in our sample.

The accuracy of the recovery of the 3D spectrum in each annulus depends on how steep the emissivity profile is. In case of the Perseus Cluster, the emissivity is strongly peaked towards the center, therefore the conversion factor significantly varies within the radial annuli. The narrower the annulus, the smaller the uncertainty. To do the conversion, we use the mean (within the annulus) value of this factor. In Section \ref{sec:uncert} we will discuss the uncertainties associated with this step of the analysis. 

\begin{figure*}
\includegraphics[trim=0 150 27 60,width=0.49\textwidth]{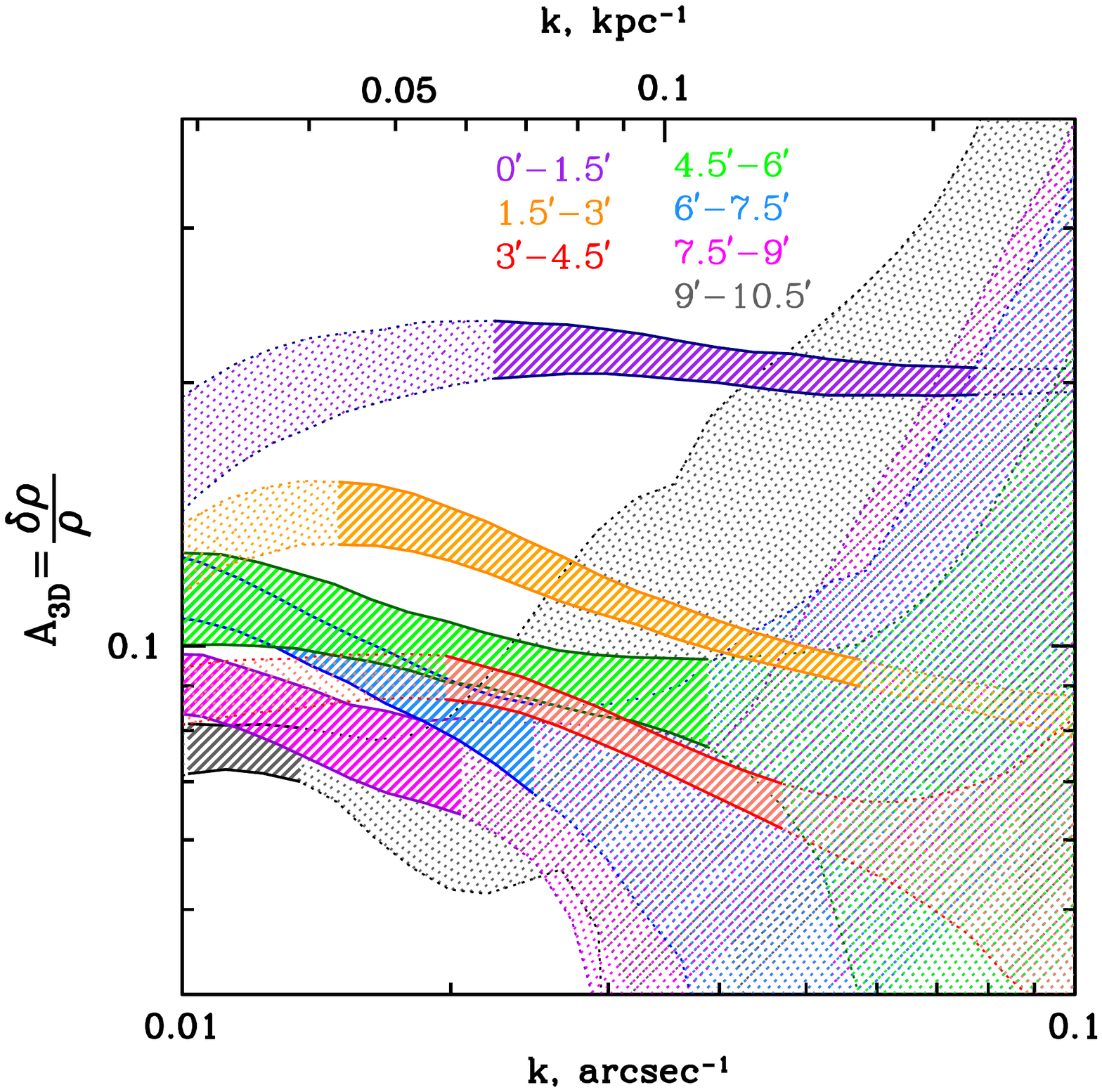}
\includegraphics[trim=-40 125 0 145,width=0.49\textwidth]{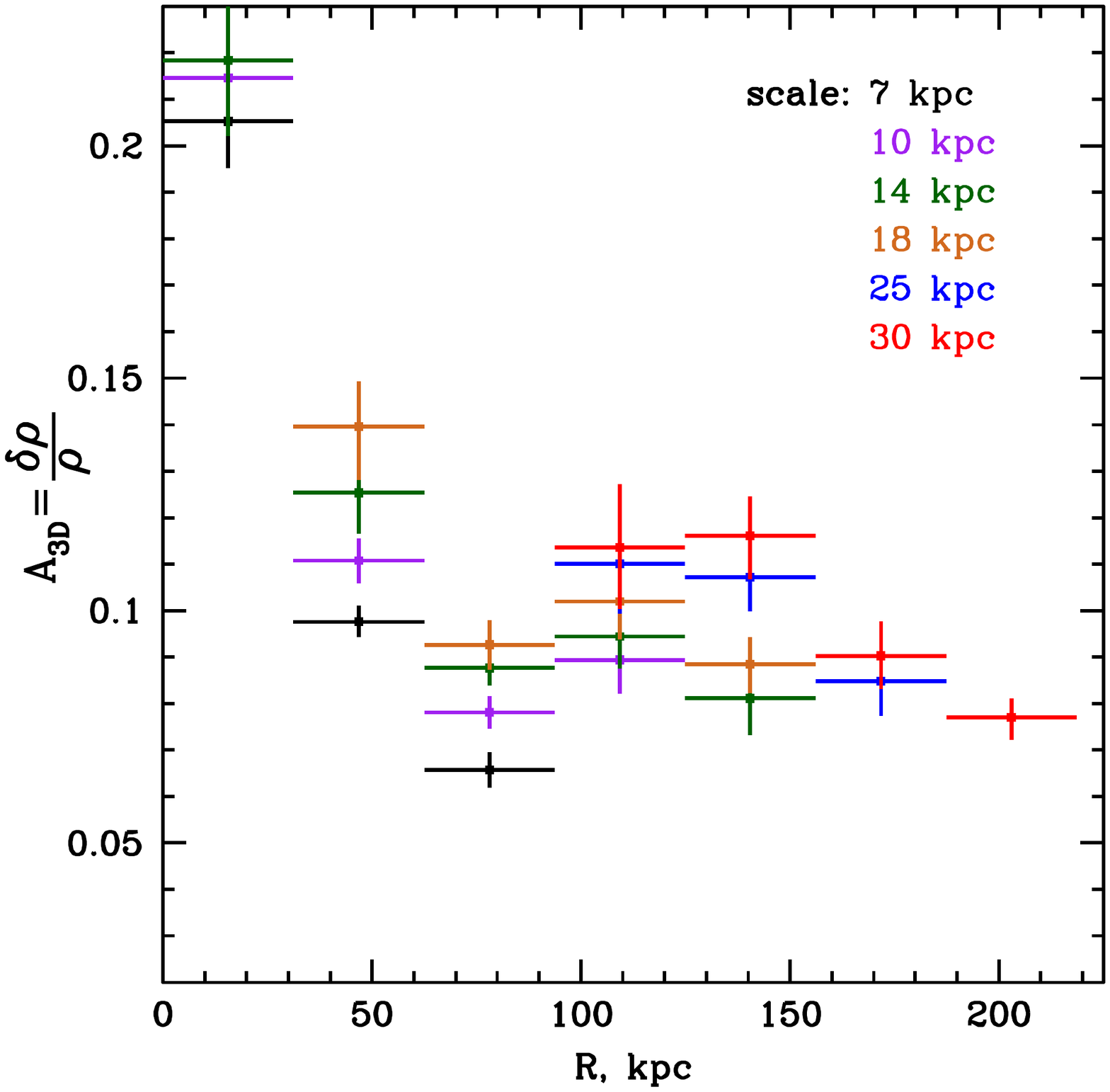}
\caption{{\bf Left:} amplitude of density fluctuations in the Perseus
  Cluster versus wavenumber calculated in a set of radial annuli shown in Fig. \ref{fig:im_pers}. Hatched regions show the amplitude on scales where the measurements are least affected by systematic and statistical uncertainties. The width of each curve reflects estimated $1\sigma$ statistical and stochastic
  uncertainties. {\bf Right:} radial profiles of the amplitude of density fluctuations measured on
  certain scales (see the legend). Notice, that the amplitude decreases
  with the radius and increases with the length scale of fluctuations. 
\label{fig:a3d_r}
}
\end{figure*}

Fig. \ref{fig:a3d_r} shows the amplitude of density fluctuations $A_{3D}$ obtained in each annulus from the PS of SB fluctuations $P_{2D}$ as a function of wavenumber $k$. Scales, which are smaller than the width of each annulus ($\approx 31$ kpc) are plotted. Hatched regions show the amplitude on wavenumbers, over which we deem the measurements are least affected by the uncertainties and more robust against the observational limitations. The choice of the high-$k$ limit is set by the statistical uncertainty or by the PSF distortions of the amplitude. For both cases, the uncertainty is less than 20 per cent in this hatched region. At low $k$, we exclude scales, on which the amplitude flattens or decreases towards smaller $k$. This flattening is most likely determined by the presence of several characteristic length scales (e.g. the distance from the center, scale heights, additional driver of fluctuations) and by the uncertainties in the choice of the underlying cluster model (see Section \ref{sec:uncert}).   

The amplitude of density fluctuations is less than $15$ per cent, except for the central $\sim 30$ kpc. Table \ref{tab:denvel} summarizes the characteristic values of the amplitude in all considered regions. The amplitude in the innermost region is noticeably higher and the curve itself is flatter than in other regions. This is not surprising, since a significant fraction of the volume of this region is dominated by sharp edges around bubbles of relativistic plasma, shocks around them, filamentary structures and absorption features, flattening the high-$k$ end of the amplitude. Simple tests showed the drop of power on small scales when these features are excluded from the analysis. The radial variations of the amplitude at different scales is shown in the right panel of Fig. \ref{fig:a3d_r}. Notice, that the amplitude decreases with the distance from the center and it increases with the scale (characteristic size) of the fluctuations. 

\begin{figure*}
\begin{minipage}{0.49\textwidth}
\includegraphics[trim=0 150 27 60,width=1\textwidth]{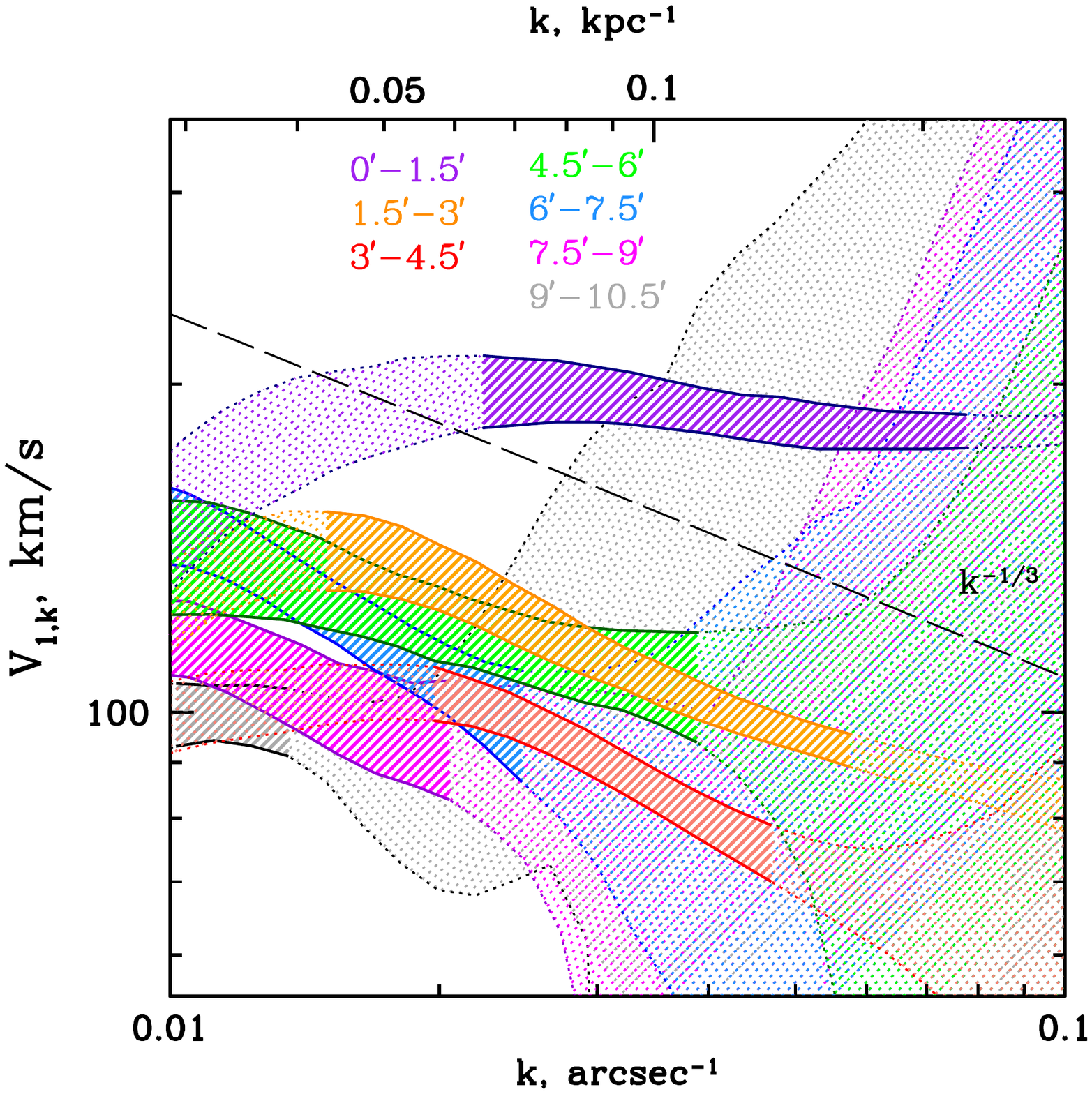}
\end{minipage}
\begin{minipage}{0.49\textwidth}
\includegraphics[trim=-45 150 0 40,width=1\textwidth]{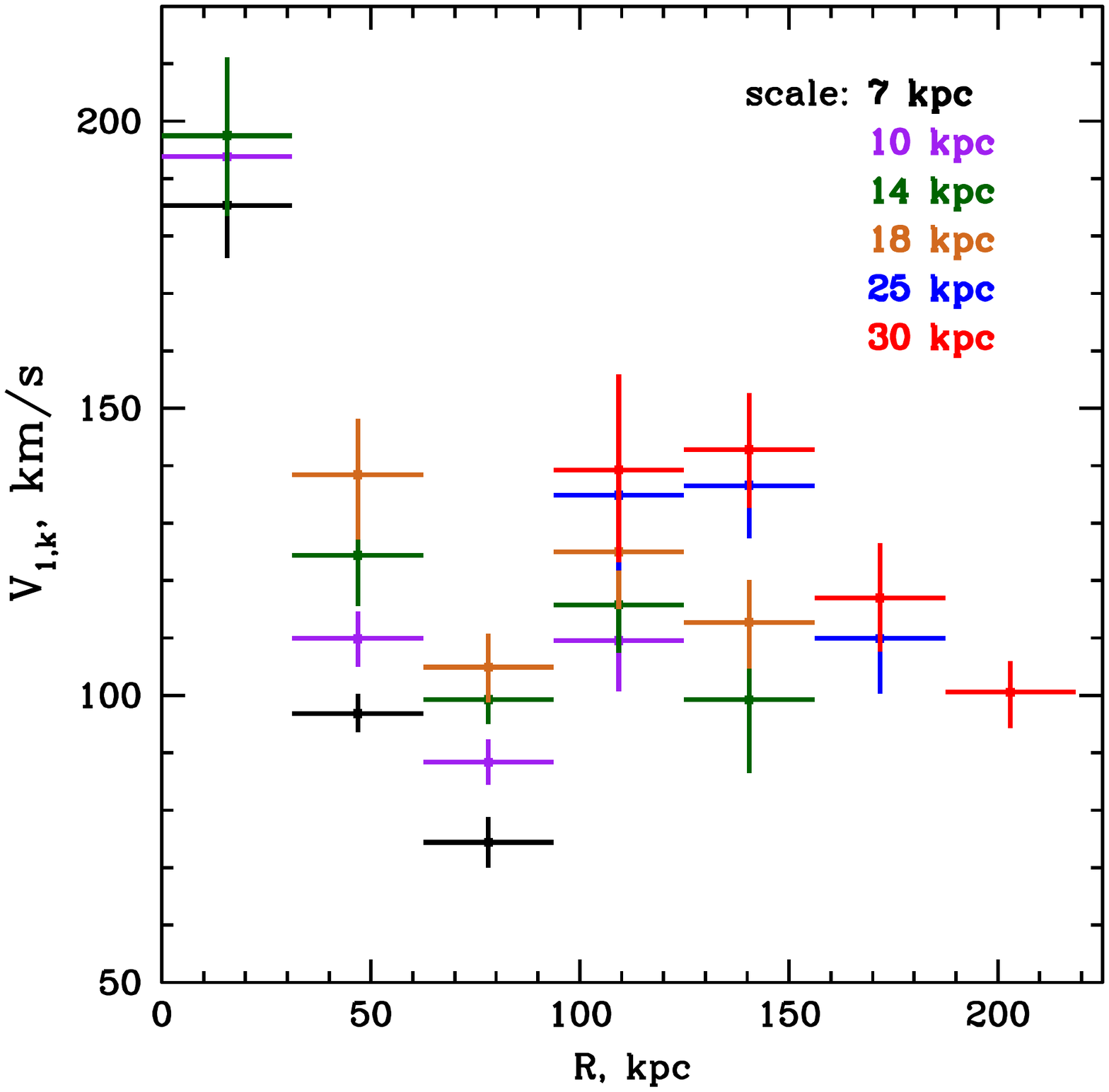}
\end{minipage}
\caption{{\bf Left:} amplitude of one-component velocity of gas
  motions versus wavenumber $k=1/l$, measured in a set of radial annuli (see the legend) in the Perseus Cluster. The velocity at each scale is obtained from the amplitude of density fluctuations, shown in Fig. \ref{fig:a3d_r}, using relation \ref{eq:rho_v}. The color-coding and notations are the same as in
  Fig. \ref{fig:a3d_r}. The slope of the amplitude for pure Kolmogorov turbulence \citep{Kol41}, $k^{-1/3}$, is shown with dash line. {\bf Right:} radial profiles of one-component velocity amplitude measured on certain length scales written in the legend.  
\label{fig:v1d_r}
}
\end{figure*}
\begin{figure}
\begin{minipage}{0.49\textwidth}
\includegraphics[trim=0 150 0 90,width=1\textwidth]{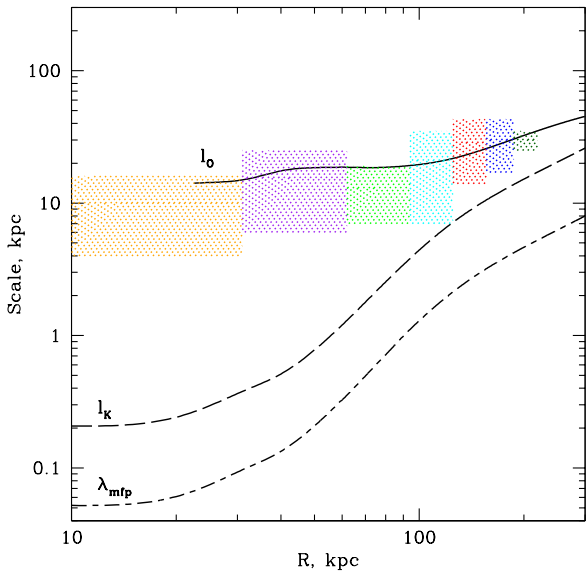}
\end{minipage}
\caption{Characteristic length scales present in the Perseus Cluster at different distances $R$ from the center. Colored shaded regions: range of scales, on which our measurements are least affected by systematic and statistical uncertainties (hatched regions in Fig. \ref{fig:a3d_r} and Fig. \ref{fig:v1d_r}). Solid curve: the Ozmidov scale $l_O$ of turbulence in the core of Perseus, obtained assuming local balance between turbulent heating and radiative cooling (relation \ref{eq:ozm}). We do not plot $l_O$ in central $\sim 20$ kpc, since the measured gas entropy is flat there, leading to $l_O\to\infty$. Dash curve: the Kolmogorov (dissipation) scale $l_K$ for unmagnetized plasma (relation \ref{eq:kolm}). Dot-dash curve: the electron mean free path for unmagnetized plasma. See Section \ref{sec:vel_ps} for discussion.
\label{fig:scales}
}
\end{figure}
\begin{table*}
\centering
\caption{The mean values of the amplitude of density fluctuations and the amplitude of one-component velocity in the Perseus Cluster (see Fig. \ref{fig:a3d_r} and \ref{fig:v1d_r}). The values are given on scales $l=1/k$: 10, 20 and 30 kpc.}
\begin{tabular}{@{}lcccccc@{}}
\hline
Annulus  & \multicolumn{3}{c}{Amplitude of density fluctuations} &  \multicolumn{3}{c}{Amplitude of one-component velocity}\\ 
         & \multicolumn{3}{c} {$A_{3D}=\delta\rho/\rho$ }        &  \multicolumn{3}{c} {$V_{1,k}$, km/s}\\
\hline
         & $l=10$ kpc & $l=20$ kpc & $l=30$ kpc & $l=10$ kpc & $l=20$ kpc & $l=30$ kpc \\
\hline
0\' \,- 1.5\' \,(0 - 31 kpc)    & 0.22 & -    & -    & 200 & -   & -\\
1.5\' \,- 3\' \,(31 - 62 kpc)   & 0.11 & 0.14 & -    & 110 & 140 & -\\
3\' \,- 4.5\' \,(62 - 94 kpc)   & 0.08 & -    & -    & 90  & -   & -\\
4.5\' \,- 6\' \,(94 - 125 kpc)  & 0.09 & 0.1  & 0.11 & 110 & 130 & 140\\
6\' \,- 7.5\' \,(125 - 156 kpc) & -    & 0.09 & 0.12 & -   & 120 & 150\\
7.5\' \,- 9\' \,(156 - 187 kpc) & -    & 0.08 & 0.09 & -   & 100  & 120\\
9\' \,- 10.5\' \,(187 - 219 kpc)& -    & -    & 0.08 & -   & -   & 100\\
\hline
\label{tab:denvel}
\end{tabular}
\end{table*}

\subsection{Velocity power spectrum}
\label{sec:vel_ps}

Knowing the amplitude of density fluctuations on each length scale, the amplitude (and the PS) of velocities of gas motions present in the ICM can be obtained. Simple theoretical arguments show that in the stratified atmospheres of relaxed galaxy clusters, the amplitude of density fluctuations $A_{3D}(k)=\disp\frac{\delta\rho_k}{\rho_0}$ and one-component velocity $V_{1,k}$ are proportional to each other at each length scale $l=1/k$ within the intertial range of scales, namely
\be
\frac{\delta\rho_k}{\rho_0}=\eta_1\frac{V_{1,k}}{c_s},
\label{eq:rho_v}
\ee
where $\rho_0$ is the mean gas density and $\eta_1$ is the proportionality coefficient \citep{Zhu14a}. If the injection scale of the turbulence is larger than or comparable with the Ozmidov scale \citep{Ozm92,Bre07}, the coefficient $\eta_1$ is set by gravity-wave physics on large, buoyancy-dominated scales, and is $\sim 1$ in atmospheres of galaxy clusters. It remains the same on smaller scales (within the inertial range) where the density becomes a passive scalar. Cosmological simulations of galaxy clusters confirm the relation \ref{eq:rho_v}, giving the averaged over a sample of relaxed clusters value $\eta_1=1\pm0.3$ \citep{Zhu14a}.

Fig. \ref{fig:v1d_r} shows the amplitude of one-component velocity versus wavenumber obtained from the relation \ref{eq:rho_v} in all seven radial annuli in the Perseus core, as well as the radial profiles of the amplitude on certain scales. Like in Fig. \ref{fig:a3d_r}, hatched regions show the amplitude on scales, where the measurements are least affected by systematic and statistical uncertainties. Notice that the velocity amplitude is higher towards the center, suggesting a power injection from the center. The spectra (amplitudes) show similar dependence on $k$, larger fluctuations are on larger $k$, consistent with cascade turbulence. The slope of the velocity PS is broadly consistent with the slope for canonical Kolmogorov turbulence, accounting for the errors and uncertainties. Within the ``robust'' range of scales, $V_{1,k}$ varies from $\sim 70$ km/s up to $\sim 210$ km/s on scales $\sim 5-30$ kpc (see Table \ref{tab:denvel}). The velocity amplitudes quantitatively match our expectations of typical velocities in the ICM from various observational constraints \citep[see e.g.][and references therein]{Chu04,Sch04,Wer09,San10,San11,deP12,San13,Pin15} and numerical simulations \citep[see e.g.][and references therein]{Nor99,Dol05,Iap08,Lau09,Vaz11,Min14}. Future direct measurements of the velocity of gas motions with X-ray calorimeter on-board {\it Astro-H} observatory \citep{Tak14} will allow us to test the density-velocity relation \ref{eq:rho_v}. Recently-proposed observational strategy for the Perseus Cluster maps the cluster core in radial and azimuthal directions, enabling us to do the calibration \citep{Kit14}.   
\begin{figure*}
\begin{minipage}{0.49\textwidth}
\includegraphics[trim=0 200 -70 150,width=1\textwidth]{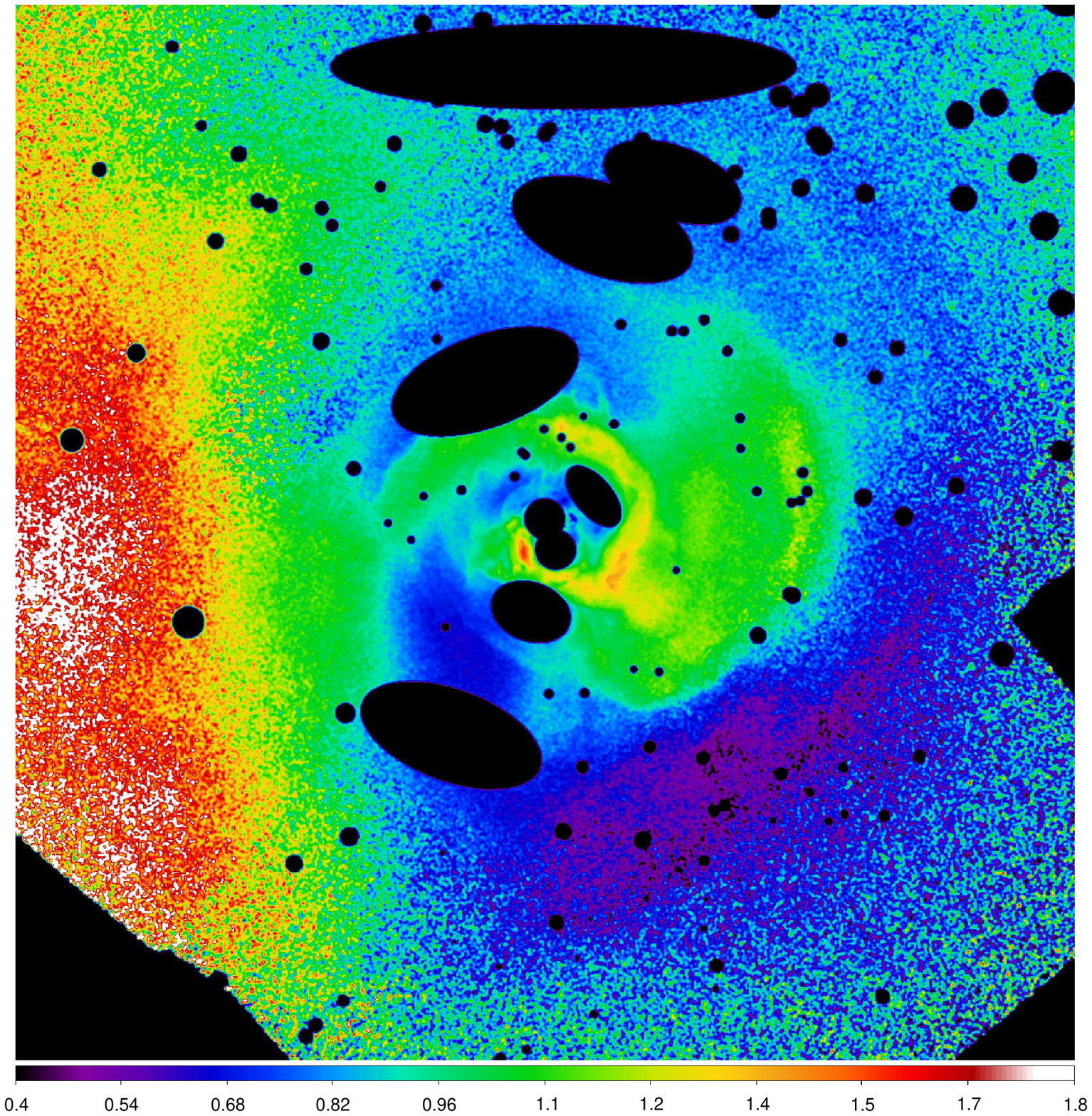}
\end{minipage}
\begin{minipage}{0.49\textwidth}
\includegraphics[trim=0 170 30 80,width=1\textwidth]{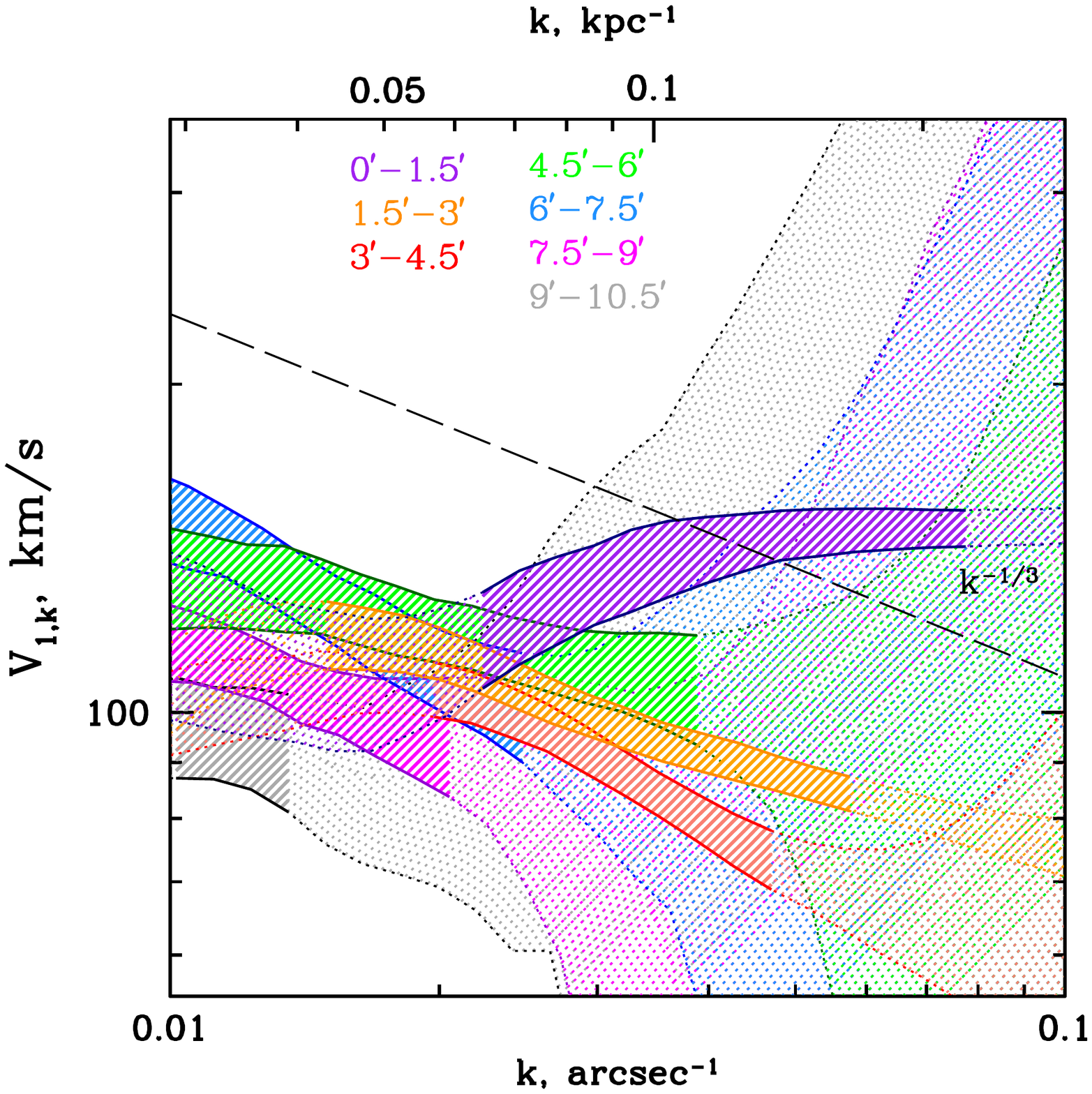}
\end{minipage}
\caption{{\bf Left:} residual image of the Perseus Cluster in the $0.5-3.5$ keV band (the same as Fig. \ref{fig:im_pers}) with excluded bubbles of relativistic plasma taken from \citet{Fab11} (black ovals). {\bf Right:} one-component velocity amplitude versus wavenumber obtained from the image on the left. The color-coding and labels are the same as in Fig. \ref{fig:v1d_r}. 
\label{fig:nobub}
}
\end{figure*}

It was recently shown that in the cores of Perseus and Virgo clusters, where the cooling time is shorter than the Hubble time, the heating of the gas due to dissipation of turbulence is sufficient to offset radiative cooling losses \citep{Zhu14b}. Assuming that the dissipation rate is matching the cooling rate, it is straightforward to estimate the Ozmidov scale $l_O$ of the turbulence using only thermodynamic properties of the Perseus Cluster. Namely,
\be
l_O=N^{-3/2}\varepsilon^{1/2}=N^{-3/2}\disp\left(\frac{Q_{cool}}{\rho}\right)^{1/2},
\label{eq:ozm}
\ee
where $N=\disp\sqrt\frac{g}{\gamma H_s}$ is the Brunt-V$\rm\ddot a$is$\rm\ddot a$l$\rm\ddot a$ frequency in the cluster atmosphere, $g$ is the acceleration of gravity, $H_s=\disp\left(\frac{d{\rm ln}S}{dr}\right)^{-1}$ is the entropy scale height, and $\varepsilon$ is the dissipation rate per unit mass. Here we assume $\rho\varepsilon\sim$ the cooling rate $Q_{cool}=n_e n_i \Lambda_n(T)$, where $n_e$ and $n_i$ are the number densities of electrons and ions, respectively, and $\Lambda_n(T)$ is the normalized gas cooling function \citep{Sut93}.

Fig. \ref{fig:scales} shows the radial profile of the Ozmidov scale $l_O$ and a range of scales we are probing in each annulus (hatched regions in Fig. \ref{fig:v1d_r}). One can see that $l_O$ is within the interval of scales we are probing at each distance from the center within the cluster core. This means that the main assumption behind the derivation in \citet{Zhu14a} is satisfied, namely the Ozmidov scale is smaller than the injection scale of turbulence (assuming that we are probing velocity PS within the inertial range). Notice, that we do not show $l_O$ at $R<20$ kpc since the measured gas entropy is flat towards the center, leading to $l_O\to\infty$.  

It is also interesting to compare the scales we are probing with the Kolmogorov (dissipation) scale 
\be
l_K=\disp\frac{\nu_{kin}^{3/4}}{(Q_{cool}/\rho)^{1/4}},
\label{eq:kolm}
\ee
where $\nu_{kin}=\disp\frac{\nu_{dyn}}{\rho}$ is the kinematic viscosity, which is obtained through the dynamic viscosity $\nu_{dyn}$ for an ionized plasma without magnetic field. Fig. \ref{fig:scales} shows the Kolmogorov scale $l_K$ as well as the mean free path for comparison. The Kolmogorov scale is significantly below the scales we are probing, which justifies even better our assumption about the inertial range of scales. In reality, the situation can be more complicated, e.g., due to the presence of magnetic fields. Here, we neglect these complications, following the simplest approach as a good starting point.

We cannot claim, of course, that the structures in the SB are due to turbulence only. For example, there are many bubbles of relativistic plasma in the Perseus core \citep[see e.g.][]{Boe93,Chu00,Fab00}, which may contribute to the signal. The question is whether their contribution is dominant in the considered regions. Various sharp structures (e.g. bubble edges) would give the slope of the amplitude $k^{-1/2}$, which is hard to discriminate from the Kolmogorov slope $k^{-1/3}$ accounting for the uncertainties of our measurements and the assumptions used. However the contribution of the bubbles to the signal can be easily seen if we repeat the analysis excluding the known bubbles from the image of Perseus. Fig. \ref{fig:nobub} shows the one-component velocity amplitude calculated from such image. Notice, that the velocity amplitude decreases by a factor $\sim 1.6-1.2$ depending on scale in the central $1.5$ arcmin. This is expected since bubbles in this region are particularly prominent and occupy a substantial fraction of the volume. In the $1.5-3$ arcmin annulus, this factor is $\sim 1.15$ over the whole range of scales. Outside $3$ arcmin, the exclusion of bubbles from the analysis does not change the measured velocity amplitude.

Sound waves \citep{Fab06,Ste09}, mergers and gas sloshing \citep{Mar07} might also contribute to the observed spectrum of the density fluctuations. Unsharp masking of the Perseus Cluster revealed quasi-spherical structures (``ripples'') in the SB, which have been interpreted as isothermal sound waves \citep{Fab00,Fab03}. An alternative interpretation is stratified turbulence, which arises naturally in the cluster atmosphere, where rough estimates give Froude number $Fr\sim 0.3-1$ \citep{Zhu14a}. In this case, the radial size of each ``ripple'' $\Delta r$ is determined by $HV/c_s$, where $H$ is the characteristic scale height and $V$ is the velocity amplitude. For example, in the $1.5-3$ arcmin annulus in Perseus, the scale height $H$ is $\sim 40-70$ kpc, the sound speed $c_s\sim 900$ km/s and the velocity measured from the SB fluctuations is $\sim 100-140$ km/s. This gives us typical radial sizes of the fluctuations $\sim 5-10$ kpc, which are consistent with those seen in the X-ray image of the cluster once the large-scale asymmetry in SB is removed. Certainly, we cannot claim that the observed fluctuations are associated with the stratified turbulence only. The question on the nature of the fluctuations will be addressed in our future work. At the very least, the measured turbulent velocities can be treated as an upper limit.

\subsection{Gas clumping}
\label{sec:clump}
\begin{figure}
\includegraphics[trim=0 170 0 80,width=0.45\textwidth]{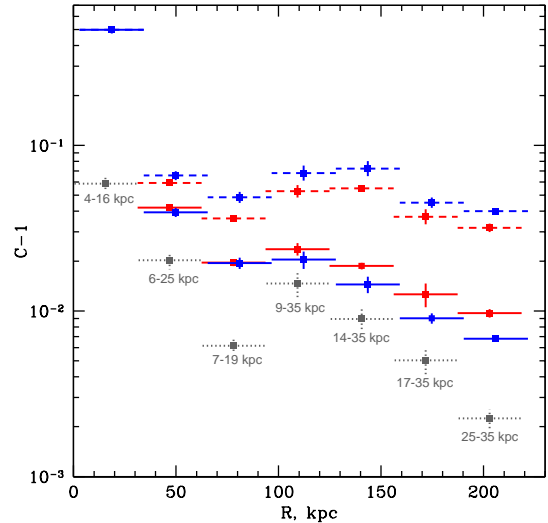}
\caption{Radial profile of the clumping factor $C-1$ in the core of the Perseus Cluster obtained from the relation (\ref{eq:clk}) using the measurements of density fluctuations power spectra. {\bf Red points}: the measured density spectrum is fitted by a power-law function with the Kolmogorov slope $-11/3$ and extended to small and large scales, which we do not probe in our analysis. {\bf Blue points:} the slope of the power-law function is assumed $-3.1$ and $-4$ for the innermost annulus $0-1.5$ and for other six regions respectively. The slope $-4$ is obtained from the power spectrum of density fluctuations in the broad annulus $1.5-10.5$ arcmin. {\bf Dash points}: $k_{min}=1/R$, $k_{max}=\infty$. {\bf Solid points:} $k_{min}=0.01$ arcmin$^{-1}$ (inverse width of each annulus), $k_{max}=\infty$. {\bf Gray dotted points:} clumping factor obtained from the measured power spectrum of density fluctuation directly, avoiding any approximations with the power-laws. Measurements within the range of scales least affected by the uncertainties (hatched regions in Fig. \ref{fig:a3d_r}) are used. The corresponding scales are written below each point. Notice, that the clumping factor is less than $7-8$ per cent outside the central $\sim 30$ kpc. 
\label{fig:clumping}
}
\end{figure}

The term ``gas clumping'' refers to any deviations of gas density isosurfaces from equipotential surfaces and, therefore, includes both large-scale inhomogeneities in the gas and small-scale clumps\footnote{Less general definitions are often used in astrophysics. For example, clumping can be referred to gas clumps and subhalos only, which are notably denser and cooler than the ambient gas.}. It is difficult to determine or even define unambiguously the surfaces of constant potential in clusters. Therefore, in practice, we measure clumping relative to a model, which describes a global gas distribution in the cluster potential well. Namely, the gas clumping factor is usually defined as
\be
C=\disp\frac{\langle (\rho/\rho_0)^2\rangle}{\langle \rho/\rho_0\rangle ^2},
\label{eq:cl}
\ee 
where $\langle\rangle$ denotes the mean or median inside a spherical shell and $\rho_0$ is the global density model, which accounts for the density gradient within each shell.  

Gas clumping leads to an overestimate of the gas density and, as a consequence, affects the gas mass and total hydrostatic mass measurements of galaxy clusters \citep[see e.g.][]{Lau09} as well as the SZ measurements of the Compton parameter \citep[e.g.][]{Khe13}. Numerical simulations predict clumping $< 1.05$ in central $0.5r_{500}$, $\sim 1.1-1.4$ at $r_{500}$ and up to $2$ at $r_{200}$ \citep[see e.g.][]{Mat99,Nag11,Ron13,Zhu13,Vaz13}. The clumping factor obtained from X-ray observations of cluster outskirts is slightly higher and varies from $\sim 2$ to $\sim 3$ at $r_{200}$ \citep[see e.g.][and references therein]{Urb14,Wal12,Mor14}. The main challenges of the clumping measurements from the X-ray observations are a very low X-ray SB in cluster outskirts, where clumping is expected to be higher, and, in contrast, a low value of the clumping factor, less than 5 per cent according to numerical simulations, in the bright cluster cores.

\begin{figure*}
\includegraphics[trim=40 270 45 250,width=1\textwidth]{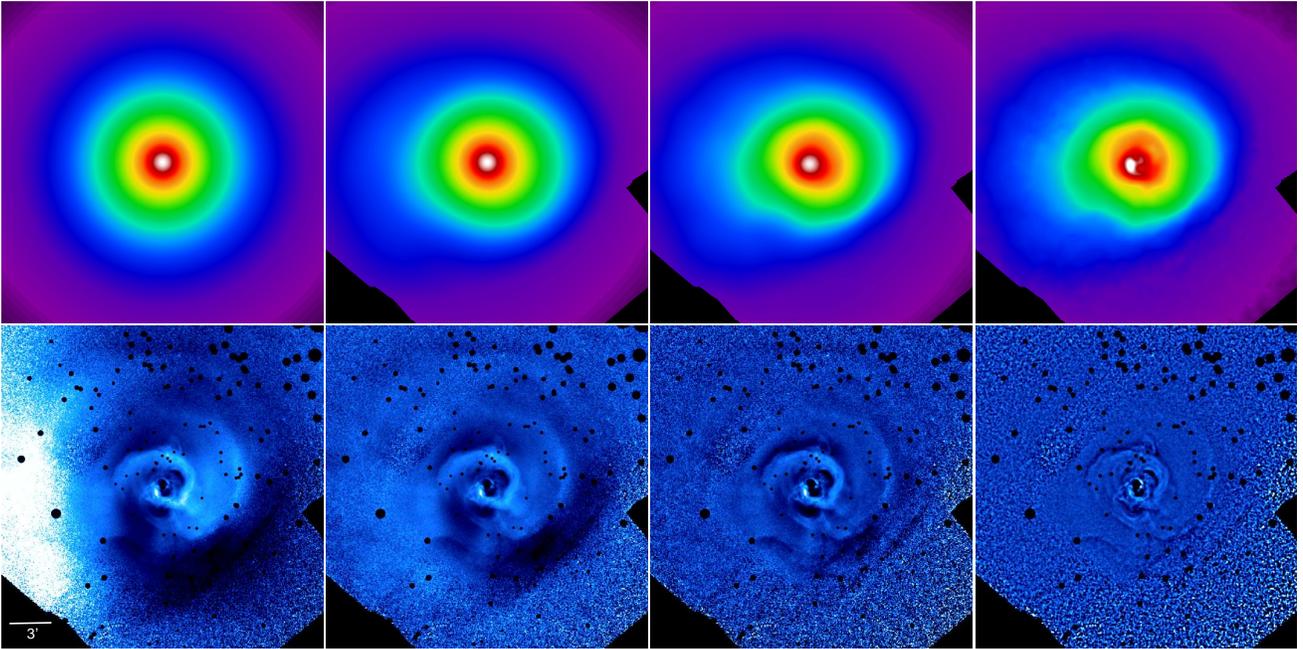}
\caption{{\bf Top:} underlying models of ``unperturbed'' surface brightness in the Perseus Cluster. From left to right: spherically-symmetrical $\beta-$model (default choice), patched $\beta-$models with $\sigma=80$ arcsec (removes large-scale asymmetry), $\sigma=30$ arcsec and $\sigma=10$ arcsec (see Section \ref{sec:und_mod} for the details). {\bf Bottom:} residual images of the SB fluctuations in Perseus obtained from the initial image divided by the underlying model (top panels). The smaller the $\sigma$ the smaller the structures included to the model and the less structures remain in the residual image.   
\label{fig:imaged}
}
\end{figure*}
Presenting the density $\rho$ through unperturbed $\rho_0$ and fluctuating $\delta\rho/\rho_0$ parts, $\rho=\rho_0(1+\delta\rho/\rho_0)$, the gas clumping factor definition (\ref{eq:cl}) can be re-written through the PS of density fluctuations as
\be
C=1+\left(\disp\frac{\delta\rho}{\rho_0}\right)^2=1+\int\limits_{k_{min}}^{k_{max}}4\pi P_{3D}(k)k^2dk,
\label{eq:clk}
\ee  
where the amplitude of density fluctuations $\disp\frac{\delta\rho}{\rho_0}$ is defined through the PS as $\disp\left(\frac{\delta\rho}{\rho_0}\right)^2=\int 4\pi P_{3D}(k)k^2dk$ and the integration is over all wavenumbers, i.e. from $k_{min}=0$ till $k_{max}=\infty$. Using this relation and our measurements of the PS of density fluctuations in the Perseus core, we estimate the gas clumping factor, which is shown in Fig. \ref{fig:clumping}. For each annulus we obtain $k_{min}$ and $k_{max}$ as the smallest and the largest wavenumbers within the range of wavenumbers least affected by uncertainties (hatched regions in Fig. \ref{fig:a3d_r}), see dotted gray points in Fig. \ref{fig:clumping}. The gas clumping is very low, less than 2 per cent at $R>\sim 30$ kpc, which is expected since we are integrating over a narrow range of wavenumbers. In order to estimate the total clumping, we approximated the PS of density fluctuations with power-law functions and extended them up to $k_{max}=\infty$ and down to $k_{min}=0.01$ arcmin$^{-1}$ (inverse width of each annulus, solid points in Fig. \ref{fig:clumping}) or down to $k_{min}=1/R$ (dash points in Fig. \ref{fig:clumping}), where $R$ is the distance from the center to each annulus. Red points in Fig. \ref{fig:clumping} show the gas clumping in case when we fit a power-law with the Kolmogorov slope $-11/3$ to the measured spectra, varying only its normalization in each annulus. For the blue points, we assume the slope being consistent with the slope of the global density spectrum obtained in the broad $1.5-10.5$ arcmin annulus. Namely, for annuli at $R>30$ kpc ($1.5$ arcmin), the slope of the spectrum is $-4$, while for the innermost annulus, the slope is $-3.1$. Notice, that the gas clumping is dominated by the large scales; it decreases with the distance from the center if we do not account for fluctuations on scales larger than the width of each annulus (it simply reflects the fact that the amplitude of fluctuations decreases with the radius, see Fig. \ref{fig:a3d_r}); outside the central $\sim 30$ kpc, the gas clumping is lower than $7-8$ per cent, which leads to a density bias $\sim (C-1)/2$ less than $3-4$ per cent.

\section{Uncertainties in the analysis}
\label{sec:uncert}
In this Section, we examine various systematic uncertainties in the analysis. Namely, we consider uncertainties associated with
\begin{itemize}
\item the choice of the unperturbed cluster density model; 
\item the bias of the $\Delta-$variance method used for the PS calculations;
\item inhomogeneous exposure coverage;
\item conversion $P_{2D}$ to $P_{3D}$; 
\item velocity measurements.
\end{itemize}
Even though the list of uncertainties is, admittedly, long, a conservative estimate of the total uncertainty on the amplitude of density fluctuations (one-component velocity) is less than $50$ ($60$) per cent.

\subsection{Underlying model of the surface brightness}
\label{sec:und_mod}

\begin{figure}
\begin{minipage}{0.49\textwidth}
\includegraphics[trim=0 170 0 90,width=1\textwidth]{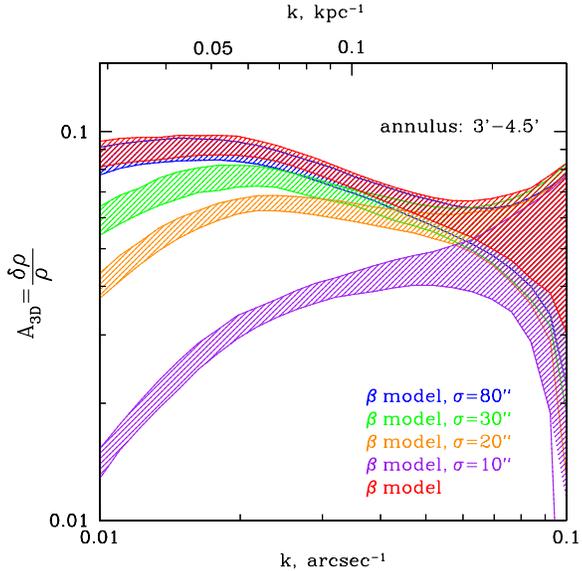}
\end{minipage}
\caption{Amplitude of density fluctuations in the $3-4.5$ arcmin annulus in the Perseus Cluster obtained by using spherically-symmetrical $\beta-$model as the underlying one (red) and more flexible models shown in Fig. \ref{fig:imaged}. Notice, that the removal of the global cluster asymmetry (blue hatched region) does not affect the amplitude measurements relative to the default model (red) on scales we are probing. Also, notice that even if the amplitude suppression is present on scales, which are included to the underlying model (as expected), the amplitude on smaller scales, which are not affected by the model, remains almost the same. This means that the measured amplitude is indeed due to the presence of fluctuations of these scales and not due to the power leakage from the large scales.
\label{fig:ampl_bmod}
}
\end{figure}

Decomposition of the cluster image into ``perturbed'' and ``unperturbed'' components is ambiguous. Physically motivated, the underlying model of the ``unperturbed'' component should reflect the global potential of the cluster in equilibrium, i.e. when the isosurfaces of the gas density and temperature are aligned with the equipotential surfaces. Our default choice of the model -  $\beta-$model for the azimuthally averaged SB (or just mean SB) - seems to be a reasonable choice of such ``unperturbed'' model at least for (or close to) relaxed galaxy clusters. However, one can see the large-scale, west - east asymmetry in the SB of the Perseus Cluster (Fig. \ref{fig:im_pers}, left), which might question our most simple and conservative choice. To check how the global cluster asymmetry affects the density amplitude measurements, we repeated the analysis, using two other $\beta-$models of the SB averaged in $180^{\circ}$ sectors. The best-fitting parameters of the $\beta-$model in the east ($90^{\circ}\div270^{\circ}$) and the west ($-90^{\circ}\div90^{\circ}$) sectors are $r_c=1.76$ and $0.89$ arcmin, $\beta=0.61$ and $0.48$, respectively. Choosing these underlying models, the maximal deviations of the density amplitude from the default case are on the largest scales in each annulus (typically $\sim 30-20$ kpc) and are less than $10$ per cent in the central 6 arcmin, less than $17$ per cent in the $6-7.5$ arcmin annulus and less than $30$ per cent at the distance $7.5-10.5$ arcmin from the cluster center.

Going beyond this simple spherically-symmetric model implies that we believe that the underlying cluster potential is more intricate. It is not clear to what degree of complexity of the model we should go. There is always a danger that some of the structures unrelated to the cluster gravitational potential are removed. We made a number of tests, introducing more flexibility to the $\beta-$model, by patching it on large scales to account for possible complexity of the potential. Namely, our patched model is defined as $I_{pm}=I_{\beta} S_{\sigma}[I_{X}/I_{\beta}]$, where $I_{\beta}$ is the $\beta-$model of azimuthally-averaged SB, $S_{\sigma}[\cdot]$ denotes Gaussian smoothing with the smoothing window size $\sigma$ and $I_X$ is the clusters X-ray SB. Varying the size of the smoothing window, the model changes from most conservative symmetrical $\beta-$model (large $\sigma$) to most complicated (and clearly implausible) one, which accounts for structures on all scales (small $\sigma$) \footnote{An alternative way to remove large-scale structure is used in the analysis of SB fluctuations in the AWM7 cluster \citep{San12}. The cluster is modeled by fitting ellipses to contours of SB, spaced logarithmically.}. Examples of such models and the corresponding residual images of the SB fluctuations are shown in Fig. \ref{fig:imaged}. Models with large $\sigma$ (e.g. $\sim 80$ arcsec) remove the large-scale (east-west) asymmetry of the cluster, while those with a smaller $\sigma$ absorb more features of the image on smaller scales. Fig. \ref{fig:ampl_bmod} shows the amplitude of density fluctuations in $3-4.5$ arcmin annulus measured from the residual images in Fig. \ref{fig:imaged}. Notice, that the removal of the large-scale asymmetry (blue hatched region) does not change the amplitude of density fluctuations relative to spherically-symmetric $\beta-$model (red hatched region) over the range of scales probed in our analysis ($< 30$ kpc). As expected, the more complex models (with the smaller $\sigma$), the broader range of scales, on which the amplitude is suppressed, and the stronger the suppression of the amplitude on the largest scales.

One can estimate the scales, on which the amplitude is expected to be suppressed depending on the size of the window function $\sigma$ used in the patched underlying models. Assuming that the PS of density fluctuations is a power-law $k^{-\alpha}$, the convolution of the image PS with the Mexican Hat filter will give \citep[see relation A8 in][]{Are12}
\be
P_{\rm{mh}}(k_r)\propto \int k^{-\alpha}\left(\frac{k}{k_r}\right)^4 e^{-2(k/k_r)^2} d^nk,
\label{eq:fil1}
\ee                  
where $k_r$ is the characteristic wavenumber. The convolution of the PS of the image, with removed large-scale part smoothed with a Gaussian, is
\be
P_{\rm{mh,\sigma}}(k_r)\propto\int k^{-\alpha}\left(\frac{k}{k_r}\right)^4e^{-2(k/k_r)^2} \left(1-e^{-2\pi ^2 k^2 \sigma^2}\right)^2 d^nk.
\label{eq:fil2}
\ee 
The ratio of both will give us an estimate of a wavenumber, on which the amplitude will be suppressed for any size $\sigma$ of the window function  used for the underlying model of the SB. For the Kolmogorov PS, $k^{-11/3}$, the suppression of the amplitude obtained using the patched $\beta-$model relative to the amplitude measured using the spherically-symmetric model is $\sim 20$ per cent on $k\approx 0.8/\sigma$. This is roughly consistent with what we see in Fig. \ref{fig:ampl_bmod}. For example, the underlying patched model with $\sigma=20$ arcsec gives a 20 per cent difference in the amplitudes of density fluctuations on $k_{\rm char}\approx0.04$ arcsec$^{-1}$. On $k<k_{\rm char}$ the ``patched'' amplitude is strongly suppressed, while on $k>k_{\rm char}$ the amplitudes remain almost the same. This means that at $k>k_{\rm char}$ the measured amplitude of density fluctuations is due to the presence of fluctuations on these scales and not due to the leakage of power from the large scales. The net conclusion is that the amplitude of density fluctuations measured on scales $< 30$ kpc is almost not affected by the choice of the underlying model, unless the cluster potential is very disturbed.     

  \subsection{Bias in the $\Delta-$variance method}
The normalization of the PS obtained through the $\Delta-$variance method may be biased slightly, depending on the slope of the PS \citep[see Appendix B in][]{Are12}. The approximations of the initial PS of SB fluctuations with a power-law functions give the slopes $\sim -3 - -3.8$ depending on the distance from the cluster center. Therefore, the normalization of the measured amplitude of density fluctuations is on average overestimated by $\sim 20-30$ per cent and slightly lower, by $\sim 10$ per cent, in the innermost ($0-1.5$ arcmin) and outermost ($9-10.5$ arcmin) annuli.  

\subsection{Inhomogeneous exposure coverage}
\label{sec:inhom_exp}
\begin{figure}
\begin{minipage}{0.49\textwidth}
\includegraphics[trim=0 160 -10 60,width=1\textwidth]{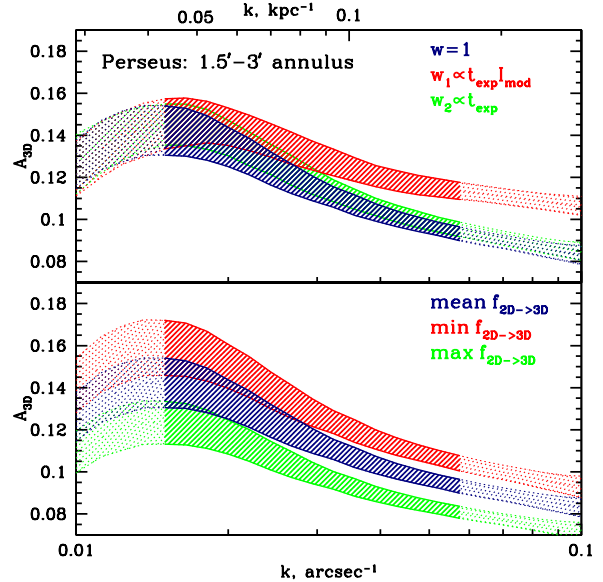}
\end{minipage}
\caption{Amplitude of density fluctuations in $1.5-3$ arcmin annulus in the Perseus Cluster measured varying the weight (top panel, see Section \ref{sec:inhom_exp}) and the conversion factor $f_{2D\to3D}$ (bottom panel, see Section \ref{sec:con2d3d}). The default choice is shown with navyblue hatched region.
\label{fig:syst_w_tw2}
}
\end{figure}

The exposure map is not uniform and the brightness of the cluster itself varies across each annulus as seen in Fig. \ref{fig:im_pers} and \ref{fig:psf_exp}. Both arguments may bias the measured amplitude of density fluctuations. By default, when calculating the amplitude by taking RMS of fluctuations present in filtered images, we use most uniform weighting scheme, $w=1$, i.e. we treat all pixels in the image with the same weight. The measured amplitude is then more sensitive to fluctuations close to the outer edge of each annulus. The least uniform scheme, but at the same time most optimal for the reduction of the Poisson noise requires weights to be $w_1\propto t_{exp}I_{mod}$, where $t_{exp}$ is the exposure map and $I_{mod}$ is the underlying model of the SB. In this case, those parts of the cluster that have higher number of counts would have larger weights. We also experimented with the weight $w_2\propto t_{exp}$, which is more sensitive to the deepest-exposure parts of the image. We find $<15$ per cent ($<30$ per cent) higher (lower) value of the amplitude of density fluctuations on scales $< 20$ kpc if weights $w_1$ and $w_2$ are used. On larger scales, $\sim 30$ kpc, the amplitude can be $< 60$ per cent lower. As an example, Fig. \ref{fig:syst_w_tw2} shows the amplitude in $1.5-3$ arcmin annulus measured using three different weighting schemes. 

\subsection{Conversion from 2D to 3D power spectrum}
\label{sec:con2d3d}
We convert the two-dimensional PS of the SB fluctuations $P_{2D}$ into the three-dimensional PS of density fluctuations $P_{3D}$ using relation (\ref{eq:p2dp3d}). The conversion factor $f_{2D\to 3D}=P_{2D}/P_{3D}$ depends on a distribution of fluctuations along the line of sight and its value varies with projected distance from the cluster center. If radial profile of the SB is steep, the uncertainty can be large unless narrow annuli are considered. We use the value of $f_{2D\to 3D}$ evaluated at the mean radius of each annulus. This factor is different in inner and outer edges of each annulus, leading to a maximal uncertainty on the measured amplitude $<17$ per cent. Fig. \ref{fig:syst_w_tw2} shows the amplitude in $1.5-3$ arcmin annulus obtained, accounting for the variations of the conversion factor $f_{2D\to 3D}$ within the annulus. 

\subsection{Uncertainties in velocity measurements}
The proportionality between the amplitude of density fluctuations and velocity with the coefficient $\eta\sim 1$ is based on the simplest approach which is a good approximation (as confirmed by numerical simulations) that captures the key physics. The derivation neglects gas heating, cooling, heat fluxes, magnetic fields and assume that all perturbations are small. Magnetic fields on large scales should not significantly modify buoyancy physics since $\beta=8\pi nkT/B^2 >>1$. On small scales, the density fluctuations would behave as a passive scalar \citep{Sch09}. Therefore, we expect the linear relation to hold even in MHD case, however the proportionality coefficient is likely to change in this case. The {\it Astro-H} X-ray observatory will allow us to verify and calibrate the relation. Any deviation from it will indicate the importance of the neglected physics. 
 
The intermittency of density fluctuations and velocity can be admittedly large in the Perseus Cluster. For example, analyzing the fluctuations in small patches within the $3-4.5$ arcmin annulus, we noticed that the amplitude of density fluctuations varies in a relatively broad range from 3 to 10 per cent. In order to achieve statistical convergence, we perform our measurements in relatively wide annuli. The amplitude obtained in the twice broader annuli is consistent with the amplitudes measured in two individual annuli.
   
\section{Conclusions}
\label{concl}
We performed detailed analysis of the X-ray SB and gas density fluctuations in a set of radial annuli within the core (central $r\sim 220$ kpc) of the Perseus Cluster, using deep {\it Chandra} observations.

To summarize our findings:
\begin{itemize}
\item The characteristic amplitude of the density fluctuations varies from $8$ to $12$ per cent on scales $\sim 10-30$ kpc within $30-160$ kpc annulus and from $ 9$ to $7$ per cent on scales $\sim 20-30$ kpc in the outer annuli, $160-220$ kpc. The amplitude in the innermost $30$ kpc is higher, up to $20-22$ per cent on scales $\sim 5-15$ kpc. The higher amplitude in this region reflects the presence of bubbles, shocks and sound waves around them, filaments and absorption feature, which occupy a large area of the considered region. The smallest scale we probe varies from $\sim 5$ kpc in the central $30$ kpc to $\sim 25$ kpc at distance $\sim 200$ kpc from the center (while the mean free path for unmagnetized plasma is $\sim 0.1$ and $5$ kpc, respectively). 
\item Given stratification and gas entropy gradient in the atmosphere of the cluster, we use linear relation between the amplitude of density fluctuations and velocity of gas motions to evaluate the characteristic velocity amplitude on different scales. The typical amplitude of the one-component velocity outside the central 30 kpc region is $\sim 90-140$ km/s on $\sim 20-30$ kpc scale and $\sim 70-100$ km/s on smaller scales $\sim 7-10$ kpc. These measurements match our expectations of typical velocities in the ICM from numerical simulations and various observational constraints. Measured velocity spectra suggest power injection from the center (e.g. from the central AGN in Perseus). Spectra are consistent with the cascade turbulence. Their slopes are broadly consistent with the slope for the canonical Kolmogorov turbulence. It was previously shown that the heating of the gas due to dissipation of such motions balances the gas radiative cooling in the cluster core \citep{Zhu14b}.
\item The gas clumping estimated from the PS of the density fluctuations is lower than $7-8$ per cent at distance from $30$ to $220$ kpc from the center, which gives a density bias less than $3-4$ per cent in the cluster core. The clumping factor is dominated by fluctuations on large scales. 
\item Systematic uncertainties in the analysis were analyzed. Conservative estimates of the final uncertainty on the amplitude of density fluctuations is $\sim 50$ per cent and on the velocity amplitude is $\sim 60$ per cent.   

Future direct measurements of the velocities of gas motions with the X-ray microcalorimeters on-board {\it Astro-H}, {\it Athena} and {\it Smart-X} will allow us to better calibrate the statistical relation between the amplitude of density fluctuations and the velocity amplitude. Any strong deviations from the proportionality coefficient $\sim 1$ would indicate the importance of the neglected physics in the ICM.  
\end{itemize} 

\section{Acknowledgements}
Support for this work was provided by the NASA through Chandra award number AR4-15013X issued by the Chandra X-ray Observatory Center, which is operated by the Smithsonian Astrophysical Observatory for and on behalf of the NASA under contract NAS8-03060. S.W.A. acknowledges support from the US Department of Energy under contract number DE-AC02-76SF00515. I.Z. and N.W. are partially supported from Suzaku grants NNX12AE05G and NNX13AI49G. P.A. acknowledges financial support from FONDECYT grant 1140304. E.C. and R.S. are partly supported by grant No. 14-22-00271 from the Russian Scientific Foundation.

\bsp
\label{lastpage}  
\end{document}